\documentclass[]{article}
\usepackage[T1]{fontenc}
\usepackage{lmodern}
\usepackage{amssymb,amsmath}
\usepackage{ifxetex,ifluatex}
\usepackage{fixltx2e} % provides \textsubscript
\IfFileExists{upquote.sty}{\usepackage{upquote}}{}
\ifnum 0\ifxetex 1\fi\ifluatex 1\fi=0 % if pdftex
  \usepackage[utf8]{inputenc}
\else % if luatex or xelatex
  \ifxetex
    \usepackage{mathspec}
    \usepackage{xltxtra,xunicode}
  \else
    \usepackage{fontspec}
  \fi
  \defaultfontfeatures{Mapping=tex-text,Scale=MatchLowercase}
  
\fi
\IfFileExists{microtype.sty}{\usepackage{microtype}}{}
\usepackage[margin=1in]{geometry}
\usepackage{longtable,booktabs}
\usepackage{graphicx}
\makeatletter
\def\ScaleIfNeeded{%
  \ifdim\Gin@nat@width>\linewidth
    \linewidth
  \else
    \Gin@nat@width
  \fi
}
\makeatother
\let\Oldincludegraphics\includegraphics
{%
 \catcode`\@=11\relax%
 \gdef\includegraphics{\@ifnextchar[{\Oldincludegraphics}{\Oldincludegraphics[width=\ScaleIfNeeded]}}%
}%
\ifxetex
  \usepackage[setpagesize=false, % page size defined by xetex
              unicode=false, % unicode breaks when used with xetex
              xetex]{hyperref}
\else
  \usepackage[unicode=true]{hyperref}
\fi
\hypersetup{breaklinks=true,
            bookmarks=true,
            pdfauthor={John M. Drake, RajReni B. Kaul, Laura Alexander, Suzanne M. O'Regan, Andrew M. Kramer, J. Tomlin Pulliam, Matthew J. Ferrari, and Andrew W. Park},
            pdftitle={Ebola cases and health system demand in Liberia},
            colorlinks=true,
            citecolor=blue,
            urlcolor=blue,
            linkcolor=magenta,
            pdfborder={0 0 0}}
\usepackage{titling}
  \title{Ebola cases and health system demand in Liberia}
  \pretitle{\vspace{\droptitle}\centering\huge}
  \posttitle{\par}
  \author{John M. Drake, RajReni B. Kaul, Laura Alexander, Suzanne M. O'Regan,
Andrew M. Kramer, J. Tomlin Pulliam, Matthew J. Ferrari, and Andrew W.
Park}
  \preauthor{\centering\large\emph}
  \postauthor{\par}
  \predate{\centering\large\emph}
  \postdate{\par}
  \date{October 30, 2014}

\begin{document}

\maketitle

\begin{abstract}
In 2014, a major epidemic of human Ebola virus disease emerged in West
Africa, where human-to-human transmission has now been been sustained
for greater than 10 months. In the summer of 2014, there was great
uncertainty about the answers to several key policy questions concerning
the path to containment. What is the relative importance of nosocomial
transmission compared with community-acquired infection? How much must
hospital capacity increase to provide care for the anticipated patient
burden? To which interventions will Ebola transmission be most
responsive? What must be done to achieve containment? In recent years,
epidemic models have been used to guide public health interventions.
But, model-based policy relies on high quality causal understanding of
transmission, including the availability of appropriate dynamic
transmission models and reliable reporting about the sequence of case
incidence for model fitting, which were lacking for this epidemic. To
investigate the range of potential transmission scenarios, we developed
a multi-type branching process model that incorporates key
heterogeneities and time-varying parameters to reflect changing human
behavior and deliberate interventions. Ensembles of this model were
evaluated at a set of parameters that were both epidemiologically
plausible and capable of reproducing the observed trajectory. Results of
this model suggest that epidemic outcome depends on both hospital
capacity and individual behavior. The model predicts that if hospital
capacity is not increased soon, then transmission may outpace the rate
of isolation and the ability to provide care for the ill, infectious,
and dying. Similarly, containment will probably require individuals to
adopt behaviors that increase the rates of case identification and
isolation and secure burial of the deceased. Given current knowledge, it
is uncertain that this epidemic will be contained even with 99\%
hospitalization rate at the currently projected hospital capacity.
\end{abstract}

\section{Introduction}\label{introduction}

The 2014 epidemic of Ebola virus in West Africa is an emerging public
health and humanitarian crisis of epic dimensions (WHO Ebola Response
Team 2014). This epidemic originated in an outbreak in Guéckédou, Guinea
in December 2013. The Ministry of Health of Guinea and Médecins Sans
Frontières (MSF) were alerted to clusters of an unknown disease with
fever/vomiting/diarrhea and a high fatality rate on March 10 and 12,
2014 (Baize et al. 2014). Through human-to-human transmission, the virus
subsequently spread to Liberia (29 March; World Health Organization
(2014e)), Sierra Leone (25 May; World Health Organization (2014f)),
Nigeria (22 July; World Health Organization (2014g)), Senegal (29
August; World Health Organization (2014h)), and the United States (30
September; World Health Organization (2014j)). On 8 August 2014 the
World Health Organization declared the epidemic to be ``a Public Health
Emergency of International Concern'' entailing an obligation on the part
of 194 signatories to participate in disease prevention, surveillance,
control, response and reporting (World Health Organization 2014i). On 6
October, the first transmission outside of Africa was documented in
Spain (Gulland 2014). As of 27 October, 13,703 persons are reported (but
not confirmed) to have been infected (World Health Organization 2014a)
with a fatality rate for those cases with known clinical outcome around
70\% (WHO Ebola Response Team 2014). Due to widespread under-reporting,
the true number of cases is widely believed to be considerably higher.

Ongoing international support has included the shipment of large
quantities of personal protective equipment, diagnostic laboratory
apparatus, and materiel such as vehicles; provision of medical and
logistical advisors from MSF, the US Centers for Disease Control \&
Prevention, and the World Health Organization, among others; and the
construction of new treatment facilities (UN-OCHA 2014). A range of
further clinical interventions, health policies, and aid are under
consideration and at various stages of mobilization. Whether these are
sufficient to achieve containment and/or what further actions might
extend their reach remain unknown. Epidemic modeling provides a means
for structured reasoning about such complex dynamical conditions, both
with respect to the information contained in this epidemic's history to
date and prospective opportunities for intervention. While several
models of the 2014 West Africa Ebola epidemic have been published, the
majority of these are primarily aimed at estimating the basic
reproduction number ($R_0$), a summary statistic that may be
tremendously informative about the potential rate of spread and the
magnitude of vaccination required to achieve herd immunity (Fisman,
Khoo, and Tuite 2014). Knowing $R_0$ is less useful where human
behaviors -- including both public health interventions (Farrar and Piot
2014) and avoidance or denial in the community (Briand et al. 2014) --
cause the epidemic to take a more irregular path. Two models that
incorporate more detail have been published. A paper by the WHO Ebola
Response Team (2014) proposes a renewal equation for the evolution of
the epidemic through time, parameterized with case reports collected by
MSF. But this model, which focuses on the time course of disease and
conditions for transmission, does not account for role of transmission
setting. The model of Meltzer et al. (2014) is more tactical, but
provides little analytical insight.

Here, we report on a model of intermediate complexity. Our goal was to
produce a model that could be used to guide policy recommendations. A
supporting objective was to perform analysis of a range of scenarios to
identify how actions taken in the present may influence short and medium
term proposects for containment. The model comprises separate
probability distributions for the number of secondary cases arising
among health care workers (HCW) infected in hospitals, non-HCW infected
by hospitalized patients, non-HCW infected during non-hospital nursing
care, and non-HCW infected through burial practices. Infected
individuals may be treated in the hospital or in the home. Hospital
treatment is assumed to result in reduced transmission but is limited to
a fixed number of available hospital beds. Cases in excess of hospital
capacity are assumed to be treated in the home. Only cases seeking
hospitalization (whether capacity allows admission or not) are scored as
a report, separating the total number of cases (which is unknown) from
the number of cases reported. In contrast to the models of WHO Ebola
Response Team (2014) and Meltzer et al. (2014), this model allows for
changing human behavior and epidemic interventions through time-varying
rates of hospitalization, exposure of health care workers, and secure
burial. We use the theory of branching processes to derive an expression
for the mean number of secondary infections.

Here we report on the application of this model to the current situation
in Liberia. We focus on Liberia for a combination of practical and
intellectual reasons. Practically, the international variation in both
transmission and response precluded treating the three countries of
Liberia, Sierra Leone, and Guinea simultaneously. Additionally, although
the situation is uniformly urgent across the three countries, the speed
of spread in Liberia during late summer and early fall of 2014 suggest
it is the country of greatest need, both with respect to the number of
persons at immediate risk of infection and as a probable regional
``driver'' of large-scale dynamics. Intellectually, we found our
modeling work to be facilitated by the consistent collection of data in
Liberia since late June with timely and disaggregated reporting, high
level coverage in the American news media, and clear plan for the
expansion of hospital capacity.

\section{Methods}\label{methods}

\subsection{Data}\label{data}

Data were obtained from situation reports issued by the World Health
Organiasation and the Liberia Ministry of Health. All situation reports
were pulled from Liberian MoH website or United Nations Office for the
Coordination of Humanitarian Affairs (UN-OCHA) provided
websites(reliefweb.int and humanitarianresponse.info). When values had
to be interpolated, data from WHO outbreak reports were used. For
provenance and reproducibility, we digitally entered our own data, which
are available from the Dryad Digital Repository at {[}URL/DOI TO BE
PROVIDED{]}. Reported cases were scored as the sum of suspected,
probable, and confirmed cases.

\begin{figure}[htbp]
\centering
\includegraphics{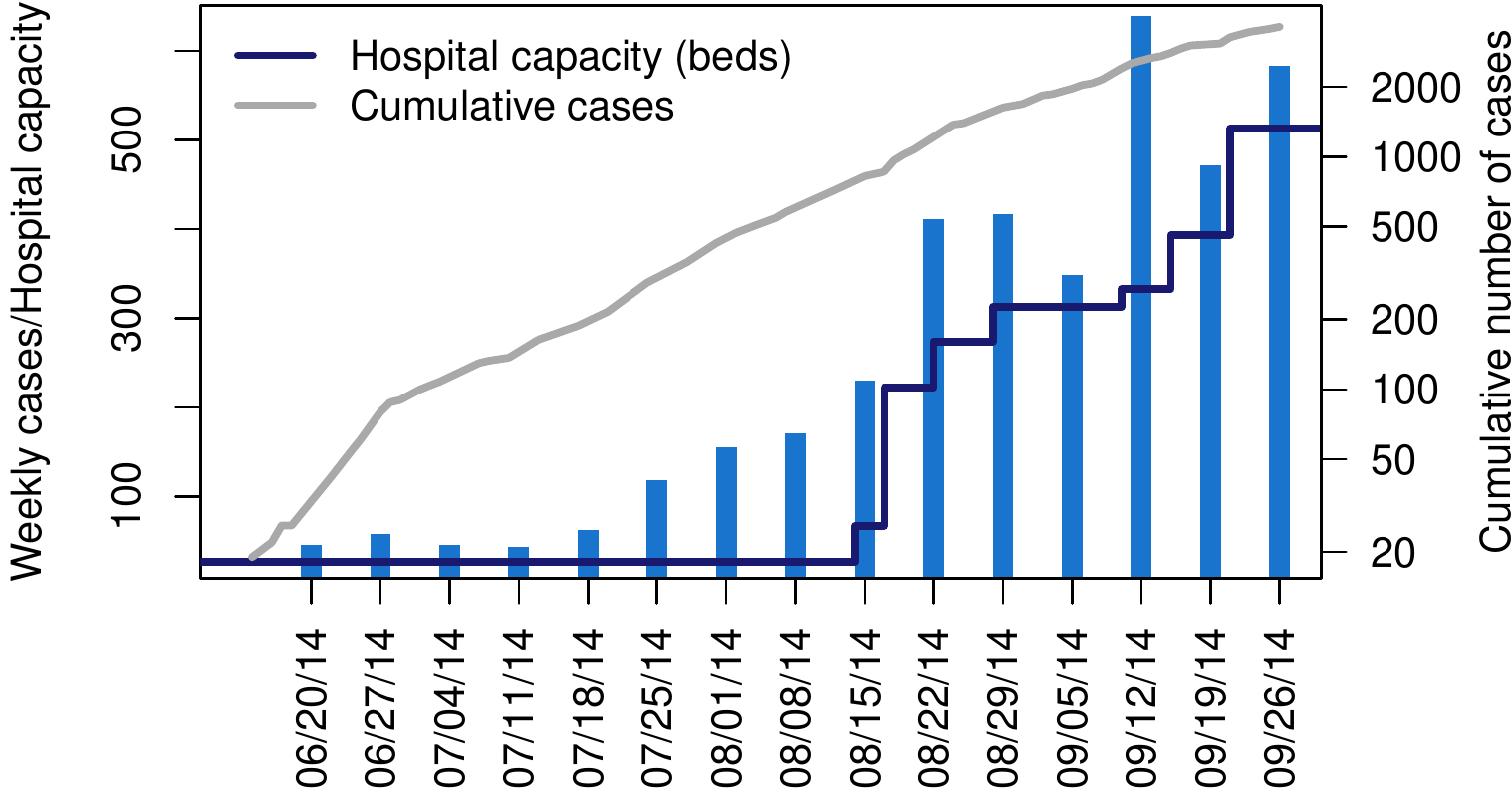}
\caption{Weekly number of suspected, probable, and confirmed cases of
Ebola virus in Liberia in the seven days terminating with each date
(blue bars) and daily cumulative reports (gray line); data from WHO
situation reports. Hospital capacity in Ebola treatment units (total
number of beds in country, dark blue line); data compiled from media and
government sources.}
\end{figure}

\subsection{A branching process model}\label{a-branching-process-model}

\subsubsection{Model}\label{model}

We developed a discrete time, stochastic process model for Ebola
transmission. The model considers where transmission occurs and who is
infected as a result. This allows a minimal set of subpopulation
differences to be articulated that nonetheless reflect the major
epidemiological properties of Ebola transmission, including hospital
treatment versus community care, transmission at funerals, and
scenario-dependent transmission risk differences during care-giving. The
model comprises separate probability distributions for the number of
secondary cases arising from (1) health care workers (HCW) infected in
hospitals, (2) non-HCW infected by hospitalized patients, (3) non-HCW
infected during non-hospital nursing care, (4) non-HCW infected through
burial practices. Infected individuals are considered to be treated
either in the hospital or in the home (Figure 2).

\begin{figure}[htbp]
\centering
\includegraphics{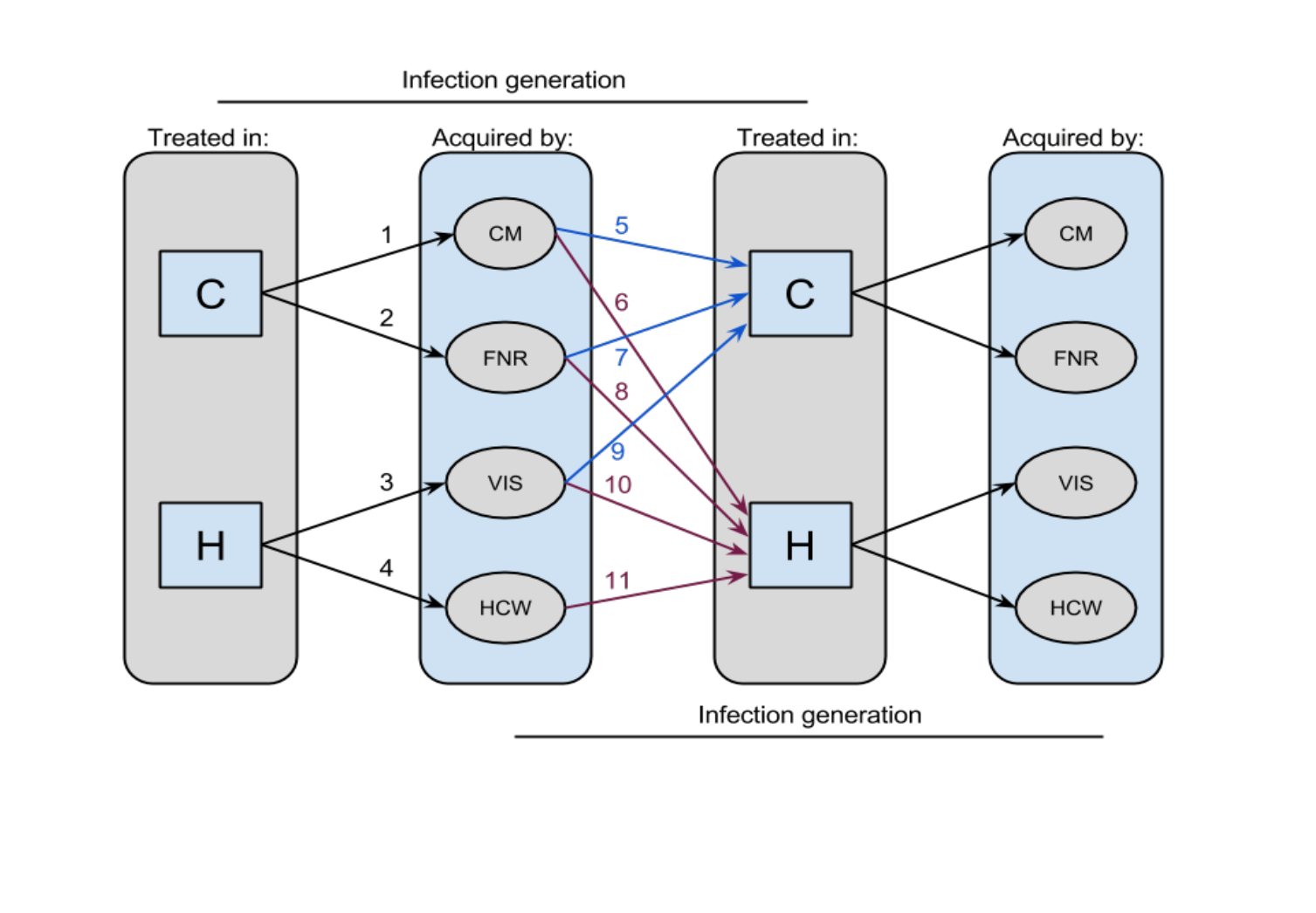}
\caption{Structure of a model for human-to-human transmission of Ebola
virus. Flow of transmission is depicted through two generations of
infection in a multi-type branching process model of Ebola virus
transmission. Grey panels show that infected persons may be treated in
either the community (C, blue paths) or hospital (H, purple paths).
Community treated patients may give rise to secondary infections in
community members through nursing care (CM) or in the process of body
preparation and burial (FNR). Hospital-treated patients may give rise to
secondary infections in health care workers (HCW) or visitors (VIS).
Infected persons may either be treated in the community or in the
hospital at rates that depend on the conditions under which the
infection was acquired.}
\end{figure}

Specifically, our model supposes that transmission is comprised of five
processes that result in 11 state transitions (Figure 2). In the
following description, numbers in parentheses correspond to labels in
Figure 2.

\begin{itemize}
\itemsep1pt\parskip0pt\parsep0pt
\item
  Persons treated in the community give rise to a Poisson distributed
  number of secondary infections among community members at rate
  $\lambda=Nq$, where $N$ is the number of contacts and $q$ is the per
  contact probability of transmission. To accommodate heterogeneity in
  transmission, $\lambda$ may be taken to be a random variable, in which
  case the number of secondary infections is negative binomially
  distributed with an additional parameter, $\theta$, regulating
  dispersion (1).
\item
  Persons treated in the hospital give rise to a Poisson distributed
  number of secondary infections among health care workers at rate
  $\lambda \beta \alpha$, where $\beta$ is a multiplier for the
  additional contacts acquired through hospitalization and $\alpha$ is a
  multiplier for the effect of infection control interventions (2).
\item
  Persons treated in the hospital may pass infection to a Poisson
  distributed number of visitors at rate $\lambda_h$ (3). It is assumed
  that all deceased hospitalized patients are given a secure burial.
\item
  Persons treated in the community recover or are given a secure burial
  at rate $g$ (and therefore do not give rise to any further secondary
  infections) and non-secure burial at rate $1-g$, giving rise to a
  Poisson distributed number of secondary infections at rate $\phi$ (4).
\item
  Persons acquiring infection in the community (classes CM, VIS, and
  FNR) are hospitalized with probability $h$ (6, 9, 11) or remain in the
  community with probability $1-h$ (5, 8, 10).
\item
  Infected health care workers are all assumed to be hospitalized (7).
\end{itemize}

These processes constitute a multi-type branching process. Branching
process models allow for very flexible specification of the distribution
of secondary cases. Branching processes do not account for the depletion
of susceptibles at the population level, however, and thus are
appropriate during the exponential phase of epidemic spread and/or where
spread is controlled through human intervention rather than
self-limitation. We believe these assumptions are broadly consistent
with the currently prevailing conditions in West Africa.

\subsubsection{Hospital capacity}\label{hospital-capacity}

In simulations, hospital treatment was assumed to result in reduced
transmission, limited by the number of available hospital beds. Patients
seeking hospitalization in excess of hospital capacity were assumed to
be returned to the home for treatment. Only patients seeking
hospitalization (whether capacity allowed admission or not) were scored
as a report, separating the total number of cases (which in reality is
unknown) from the number of cases reported.

\subsection{Parameterization}\label{parameterization}

To parameterize this model, we were initially guided by reports on the
outbreaks of Ebola virus in Kikwit (Democratic Republic of Congo) in
1995 (Bwaka et al. 1999) and Gulu (Uganda) in 2000-2001 (Oyok 2001).

\subsubsection{Transmission ($N$, $q$, $\theta$) and the effectiveness
of infection control
($\alpha$)}\label{transmission-n-q-theta-and-the-effectiveness-of-infection-control-alpha}

The attack rate in Kikwit was 9\% among hospital workers (Tomori et al.
1999) and 16\% among family members (Dowell et al. 1999). The ratio of
exposures to index cases in households was $\hat N=173/27=6.4$ for 27
different families. Assuming exposure was only within the family (so
each secondary case had only one exposure), we have $\hat q = 0.16$
(risk of transmission per contact). At Kikwit General Hospital, 37 of
429 workers met the case definition for Ebola virus disease. A reported
three cases occurred after the use of barrier nursing. If we assume that
these 3 were all in Kikwit General Hospital, then 34 health care workers
were infected prior to infection control. 110 out of 138 other hospital
workers reported direct contact with an Ebola patient. Extrapolating to
the 392 health care workers who weren't infected, we estimate the number
of workers with direct contact to be
$110/138 \times 392 + 34 \approx 346$ yielding an attack rate of 9.8\%.
Of course, hospital workers experience greater exposure than persons
providing care in the community. Among 48 uninfected persons with direct
contact jobs at Kikwit General Hospital there were a total of 151
patient contacts (3.15 contacts per worker). If this were
representative, then we would have the relation
$1-(1-q \alpha)^{3.15} = 0.098$, yielding $\alpha = 0.20$ prior to the
implementation of barrier nursing and other infection control measures.
Following barrier nursing, 3 out of $110/138 \times 392 + 3 \approx 315$
health care workers were infected, yielding an attack rate of 0.95\%.
Using the relation $1-(1- q \alpha)^{3.15} = 0.0095$ we obtain
$\alpha = 0.019$ after the implementation of barrier nursing and other
infection control measures.

\subsubsection{Hospital contact multiplier
($\beta$)}\label{hospital-contact-multiplier-beta}

The parameter $\beta$ relates the number of contacts in a health
facility to those in a household and is expressed as a multiplier of
$N$. This value is chosen based on intuition and narrative reports. In
general, we consider values in the range $2 < \beta < 5$ to be
reasonable.

\subsubsection{Funeral transmission
($\phi$)}\label{funeral-transmission-phi}

Legrand et al. (2007) assumed that mean duration of death to burial was
2 days and estimated transmission rates of 7.66 per week (Kikwit) and
0.46 per week (Gulu). Translating into average number of infections, the
number of secondary cases through funeral are estimated to be 2.18 and
0.13, respectively, assuming $S/N \approx 1$, where $S$ is the number of
susceptible individuals in the population and $N$ is the total
population size. A value of $0 <\phi<3$ is consistent with the routine
finding that preparation of the body constitutes a substantial risk
factor and that this duty is performed by a relatively small number of
people. We note that this is not consistent with anecdotal reports of
large numbers of persons being infected at a funeral. We consider those
events most likely to be exceptional. Parameter values of this ``core
model''" are reported in Table 1.

\begin{longtable}[c]{@{}lc@{}}
\toprule\addlinespace
Variable & Value
\\\addlinespace
\midrule\endhead
Household contacts ($N$) & 6.4
\\\addlinespace
Transmission probability ($q$) & 0.16
\\\addlinespace
Overdispersion ($\theta$) & 1
\\\addlinespace
Hospital contact multiplier ($\beta$) & 4
\\\addlinespace
Effectiveness of infection control ($\alpha$) & 0.019
\\\addlinespace
Average number of secondary community cases from hospitalized patients
($\lambda_h$) & 0.3
\\\addlinespace
Average number of secondary cases from a funeral ($\phi$) & 2.18
\\\addlinespace
\bottomrule
\end{longtable}

Table 1. Parameter values of the basic branching process model for Ebola
transmission.

\subsubsection{Treatment facilities}\label{treatment-facilities}

From a range of reports, we compiled a time series of the operational
Ebola Treatment Units (ETUs) along with estimates of their capacity,
recorded as the number of patient beds available (Figure 1).
Importantly, many ETUs were regularly reported to be operating above
capacity, typically by around a factor of two (World Health Organization
2014b, Medecins Sans Frontieres (2014), World Health Organization
(2014c)). Additionally, the average hospital stay is around 6.5 days
(WHO Ebola Response Team 2014), considerably shorter than the 15 day
infection generation. Therefore, throughout our analysis, we estimate
the number of patients potentially served by an ETU within an infection
interval using the formula

\begin{equation}
s(t) = 2 b(t)\tau/\sigma
\end{equation}

where $t$ marks time in infection generations, $b(t)$ is hospital
capacity in terms of the number of beds, $\tau=15$ is infection
generation time, and $\sigma=6.5$ is the average duration of
hospitalization.

\subsubsection{Secure burial rate}\label{secure-burial-rate}

Non-secure burial (including body preparation and funeral ceremonies) is
one of the key occasions for Ebola virus transmision. The Liberia
Ministry of Health and international partners have therefore sought to
reduce this mode of transmission through public education about the risk
of exposure from deceased Ebola patients and the mobilization of body
retrieval and burial teams. It is likely, therefore, that there has
already been a reduction in transmission due to increased frequency of
secure burial. For example, even during the interval from 4 July to 2
September (prior to a potentially spurious downturn (World Health
Organization 2014a)), the cumulative reported number of cases shows a
negative curvature on a logarithmic scale (grey line in Figure 1). We
therefore modeled $g$ (a rate that is the sum of the recovery rate and
secure burial rate) using the time-dependent function

\begin{equation}
g(t) = \gamma_1 (1-1/((t-7)\gamma_2)) + \mu
\end{equation}

where $t$ is measured in terms of infection generations and $\mu=0.3$ is
one minus the case fatality rate, $\gamma_1 < 0.7$ is the maximal secure
burial rate, and $\gamma_2$ governs the speed at which safe burials
increase. This function allows for the secure burial rate to increase
beginning around 4 July, starts at a positive minimum due to natural
recovery, and asymptotically approaches a maximum at $\gamma_1 + \mu$,
since we suppose that secure burial and recovery cannot go to 100\%.

\subsubsection{Initial conditions}\label{initial-conditions}

According to our data, there were 27 beds in ETUs in Liberia on 4 July.
Based on the reported cumulative case count, there were approximately
$108-30=78$ active reported cases at this time. Using equation (1), we
estimated that 54 of the reported cases were under hospital care.
Further assuming under-reporting by a factor of 2.5 (Meltzer et al.
2014, Chan (2014)), we estimated that there were a total of 195 cases
for $195-78=117$ unreported cases at this time. Together, these
calculations imply that 54 persons were treated in hospitals and 141
persons were treated in the community.

\subsubsection{Other parameters}\label{other-parameters}

In general, we treat the hospitalization rate ($h$), secure burial rate
($\gamma_1$ and $\gamma_2$), funeral transmission $\phi$, and
overdispersion ($\theta$) as tuning parameters. The time scale of this
model is defined with respect to infection generations. To calibrate to
calendar time, we assumed a serial interval of 15 days (WHO Ebola
Response Team 2014). To calculate hospital capacity, we assumed an
average hospital stay of 6.5 days (WHO Ebola Response Team 2014).

\subsubsection{Plausible parameter sets}\label{plausible-parameter-sets}

Guided by these crude parameter estimates, we then tuned our model to
data from the 2014 Ebola outbreak in Liberia. There were two waves of
transmission in 2014 in Liberia. The first wave occurred in March and
April, comprised a total of 8 reported cases, and may have gone extinct
in mid-May. The second wave began in late May and was the origin of the
vast majority of cases. However, reported cases between the end of the
first wave and around 4 July were irregular, whereas after 4 July there
was a dramatic and sustained increase in the number of cases for many
weeks. Around 6 September, the smoothed average number of cases per case
(a model-independent estimate of $R_{eff}$) began to decline (not
shown). The World Health Organization Situation Report of 8 October
indicates that this decline was probably due to a deterioration in
reporting, rather than a true decline in transmission. For these
reasons, we focused our fitting on the interval from 4 July 2014 to 2
September 2014. In keeping with the time scale of our model, and to
smooth over daily variations in reporting, reported cases were
aggregated to 15 day transmission generations (Table 2).

\begin{longtable}[c]{@{}lll@{}}
\toprule\addlinespace
Date & Cumulative cases & Cumulative cases among health care workers
\\\addlinespace
\midrule\endhead
7/4/2014 & 122 & 12
\\\addlinespace
7/19/2014 & 197 & 20
\\\addlinespace
8/3/2014 & 498 & 64
\\\addlinespace
8/18/2014 & 972 & 115
\\\addlinespace
9/2/2014 & 1847 & 153
\\\addlinespace
\bottomrule
\end{longtable}

Table 2. Reported cases and reported cases among health care workers
during five infection generations of the 2014 outbreak of Ebola in West
Africa.

The parameters $h$, $\gamma_1$, $\gamma_2$, $\theta$, $\alpha$ and
$\phi$ were first tuned so that the median simulated reports of
infection among health care workers in the four infection generations
between 4 July and 2 September and the median simulated number of
cumulative reports among non-healthcare workers at the same times were
as close to the reported values as possible (see Online Supplement). We
further refined these fits by minimizing squared differences on a
logarithmic scale. We then used latin hypercube sampling to explore a
parameter space within +/- 25\% of the tuned values. A parameter set was
deemed plausible if the reported cumulative number of cases and reported
cumulative cases among health care workers were within the observed
range of 500 simulations.

\subsection{Forecasting}\label{forecasting}

To forecast future cases under different scenarios for aid and
intervention, we project cases and number of persons seeking
hospitalization from 3 September 2014 until 31 December 2014 (120 days)
under five scenarios:

\begin{enumerate}
\def\labelenumi{\arabic{enumi}.}
\itemsep1pt\parskip0pt\parsep0pt
\item
  \emph{Baseline.} Transmission and hospitalization continue at current
  levels (hospital capacity of 601 beds);
\item
  \emph{Scenario A.} Conditions improve due to the U.S. aid commitment
  of 15 September 2014 (hospital capacity increases by 1,700 beds in
  Ebola treatment centers between 25 October 2014 and 28 December 2014
  to a total of 2,301 beds);
\item
  \emph{Scenario B.} Conditions improve through an increase in hospital
  capacity of 6,800 new beds (four times the U.S. aid commitment of
  1,700 beds), bringing total hospital capacity to 7,401 bed equivalents
  by 28 December 2014;
\item
  \emph{Scenario C.} Conditions improve by increase in hospital capacity
  to 7,401 bed equivalents by 28 December 2014 and hospital admission
  rate of 85\%.
\item
  \emph{Scenario D.} Conditions improve by increase in hospital capacity
  to 7,401 bed equivalents by 28 December 2014 and hospital admission
  rate of 99\%.
\end{enumerate}

Initial conditions for these scenarios were derived from outbreak
reports issued by the Liberia Ministry of Health and World Health
Organization. Specifically, on 2 September 2014 the number of persons
per infection generation that could be treated in ETUs was 1444. In this
generation, the number of reported infected persons was $1871-972=899$
for a total infection generation of approximately 2248. We assume that,
at most, the fraction seeking hospitalization (60.2\%) was admitted,
yielding $2248 \times 0.62 \approx 1394$ with $2248-1394 = 854$
remaining in the community.

\section{Results}\label{results}

\subsection{Model fit}\label{model-fit}

Overall, 1,045 of 5,000 (20.9\%) parameter sets were determined to be
plausible. Mean values from plausible parameter sets are reported in
Table 3.

\begin{longtable}[c]{@{}lc@{}}
\toprule\addlinespace
Variable & Value
\\\addlinespace
\midrule\endhead
Hospitalization rate ($h$) & 0.6
\\\addlinespace
Overdispersion ($\theta$) & 2.2
\\\addlinespace
Average number of secondary cases from a funeral ($\phi$) & 5.9
\\\addlinespace
Speed of secure burial improvement ($\gamma_0$) & 0.6
\\\addlinespace
Average number of secondary cases from a funeral ($\phi$) & 6.0
\\\addlinespace
Core secondary transmission rate ($\lambda$) & 1.1
\\\addlinespace
Hospital leakage ($\lambda_h$) & 0.25
\\\addlinespace
\bottomrule
\end{longtable}

Table 3. Mean values of plausible parameter sets.

The fit of the tuned model to the cumulative number of reported cases in
Liberia is shown in Figure 4. The heavy blue line shows the cumulative
number of reported cases. The plausible range of case reports given the
model is shown in yellow (95\% prediction intervals). The plausible
range of total cases, including unreported cases, is shown in blue. The
fit of the model to infection generations in health care workers and in
the general public is shown in Figure 5.

\begin{figure}[htbp]
\centering
\includegraphics{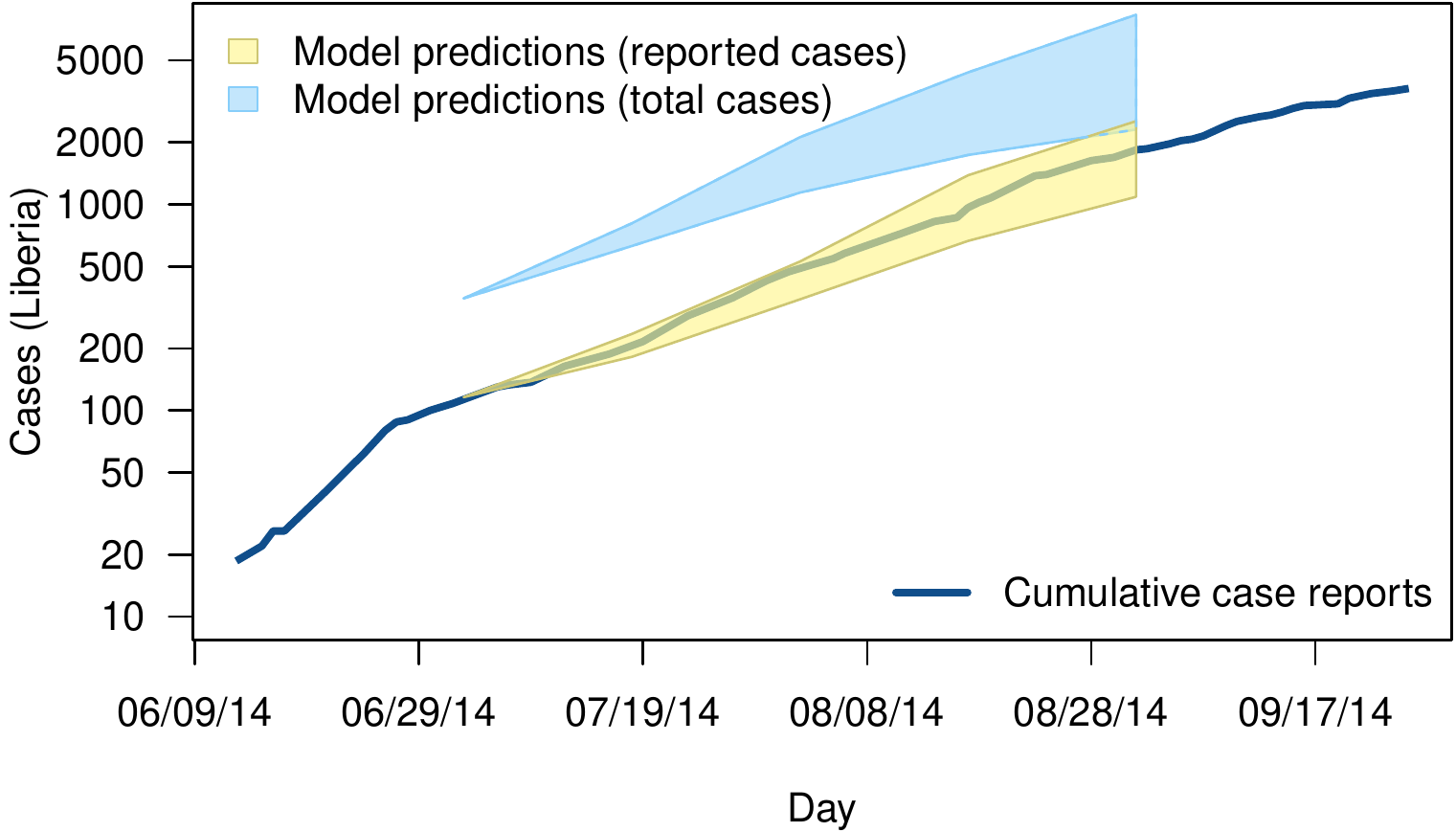}
\caption{The fit model for Ebola transmission in Liberia initialized on
4 July. The heavy blue line shows the cumulative number of cases
reported. The yellow region shows the model-predicted range of cases
expected to be reported given incomplete reporting. The blue region
shows the model-predicted total number of cases over the same time.}
\end{figure}

\begin{figure}[htbp]
\centering
\includegraphics{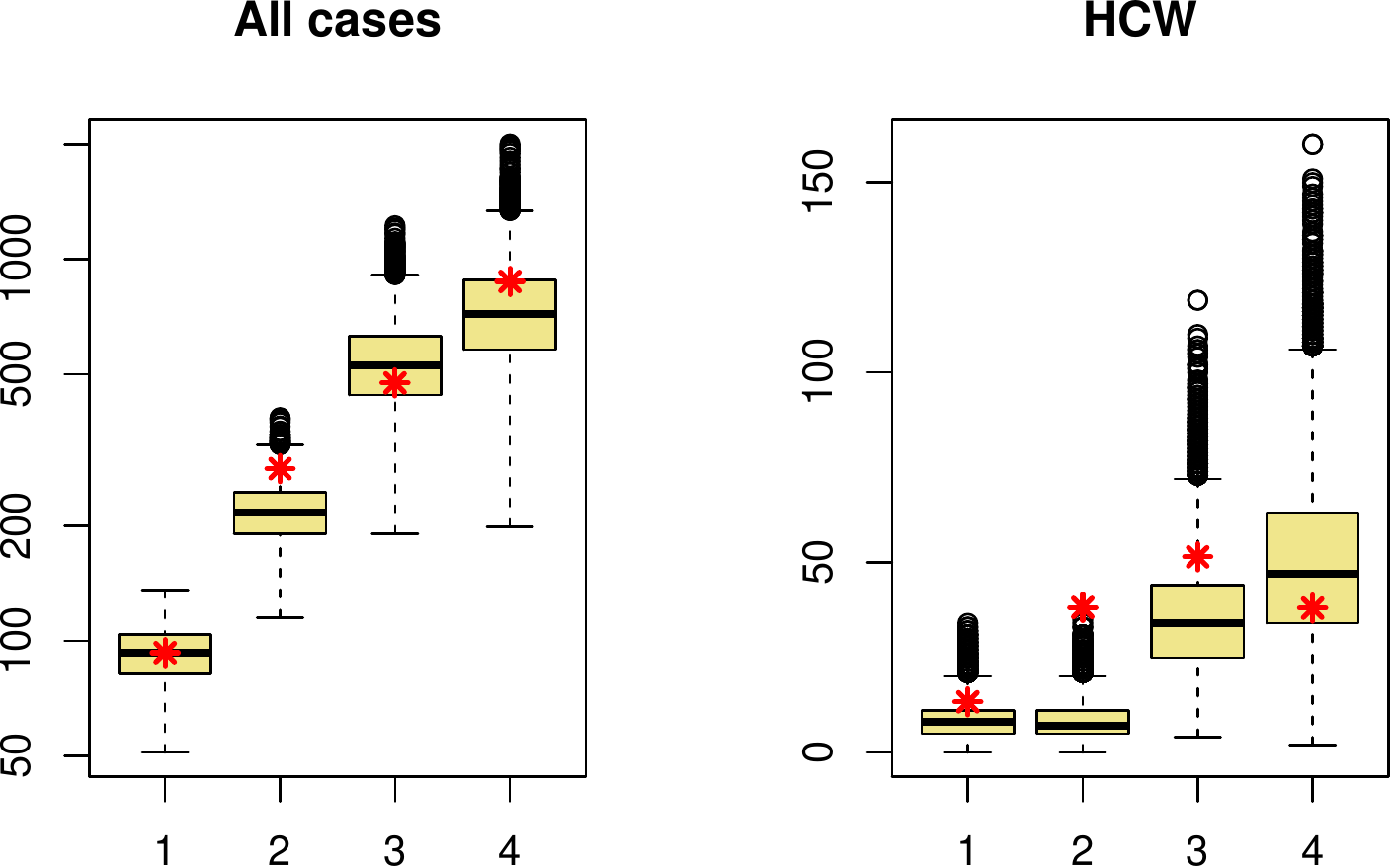}
\caption{Comparison between the total number of reported cases (red
asterisks) and model-generated distribution (box-and-whisker plots)
during four generations of the epidemic, starting 4 July 2014. Shown are
values for all cases (left panel) and cases among health care workers
(right panel).}
\end{figure}

\subsubsection{Effective reproduction
number}\label{effective-reproduction-number}

Model-based effective reproduction numbers at infection generations
between 4 July and 17 October were calculated by evaluating the
effective reproduction number (see online supplement) at the 1,045
plausible parameter sets. The change over time in the range of plausible
effective reproduction numbers is shown in Figure 3.

\begin{figure}[htbp]
\centering
\includegraphics{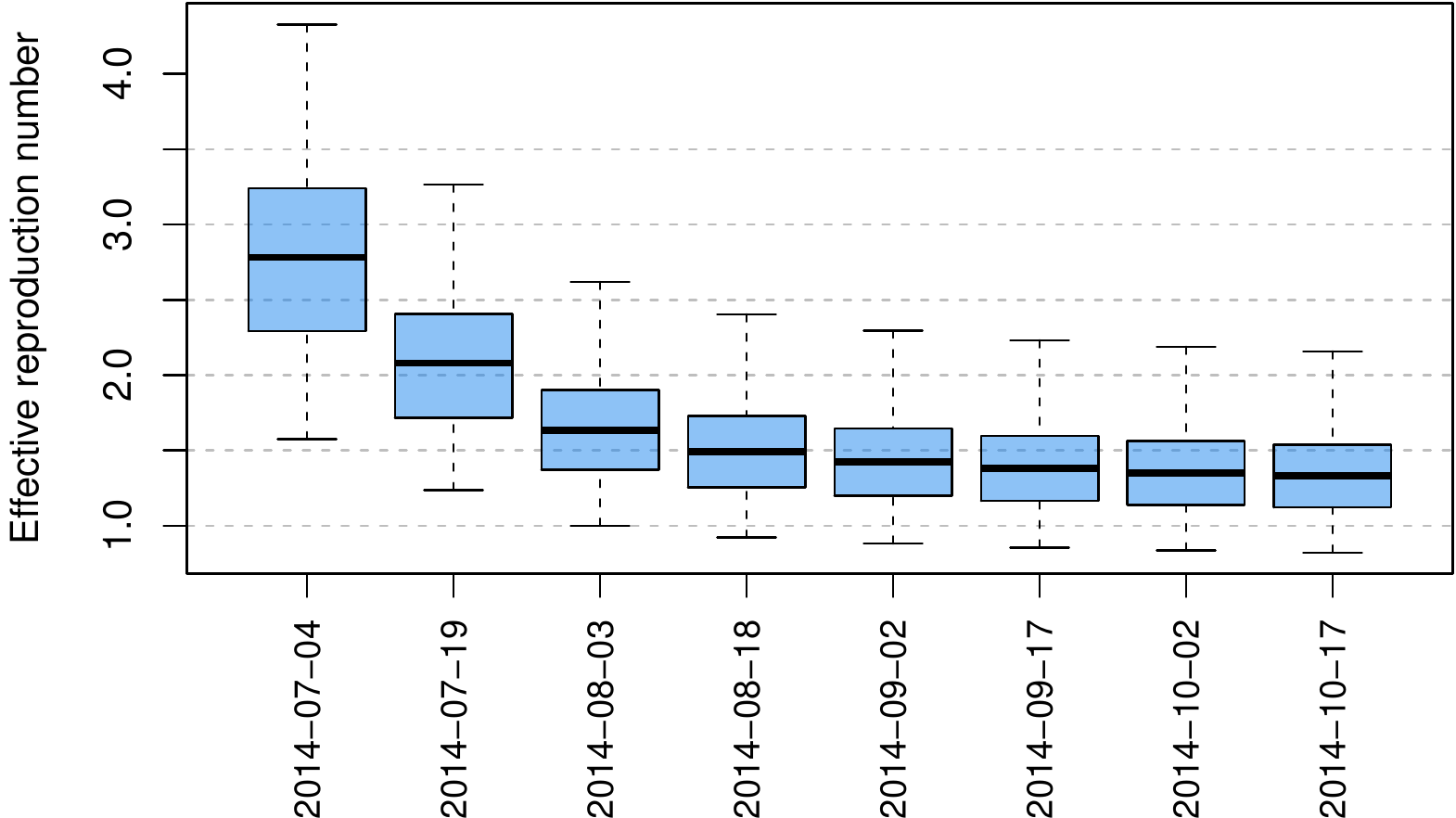}
\caption{Effective reproduction numbers for Ebola virus in Liberia from
July-October 2014.}
\end{figure}

\subsection{Forecasts and containment}\label{forecasts-and-containment}

Simulated trajectories illustrating the possible outcomes starting on 2
September assuming baseline conditions are shown in Figure 6. The median
projected total epidemic size by 31 December is 130,862 cases
(inter-quartile range: 44,560 to 396,706). The top panel shows the range
of trajectories for $10,450$ simulations distributed over 1,045
plausible parameter sets. An interpolation to project the daily number
of persons seeking hospitalization is contained in the online
supplementary materials. Projected epidemic size by 31 December for all
five scenarios is shown in Figure 7.

\begin{figure}[htbp]
\centering
\includegraphics{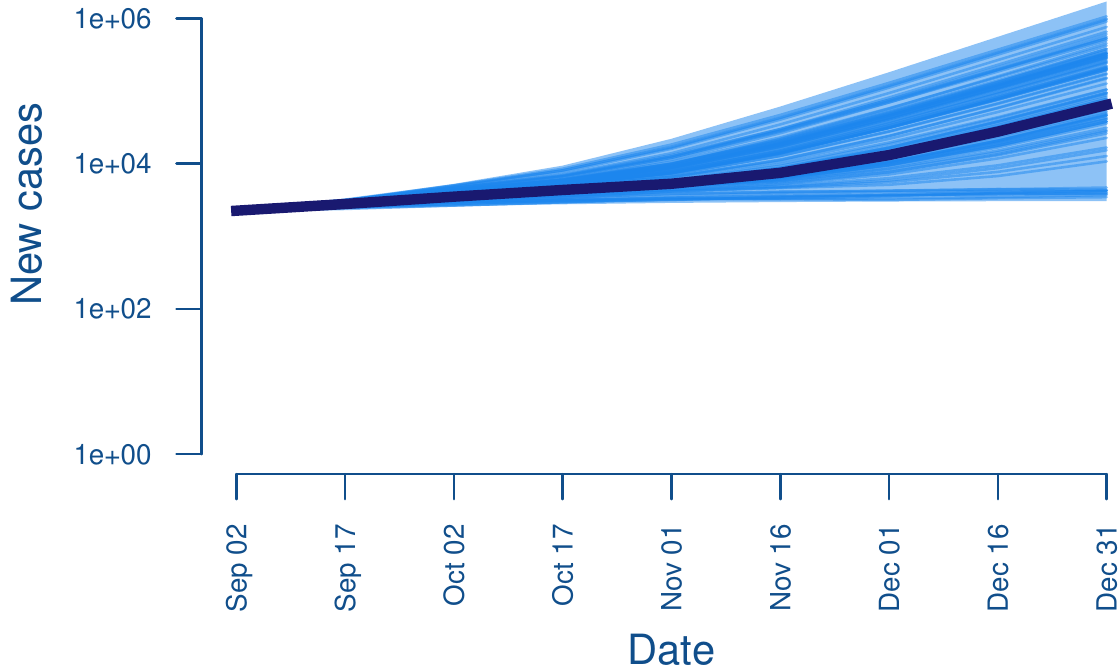}
\caption{Number of cases in each infection generation when transmission
occurs at the baseline rate. Dark blue lines show 100 stochastic
realizations of the model. Shaded region shows 95\% quantiles over
$10,450$ realizations; x-axis is the final date of the infection
generation, e.g., 17 September is for all cases in the infection
generation beginning 2 September.}
\end{figure}

\begin{figure}[htbp]
\centering
\includegraphics{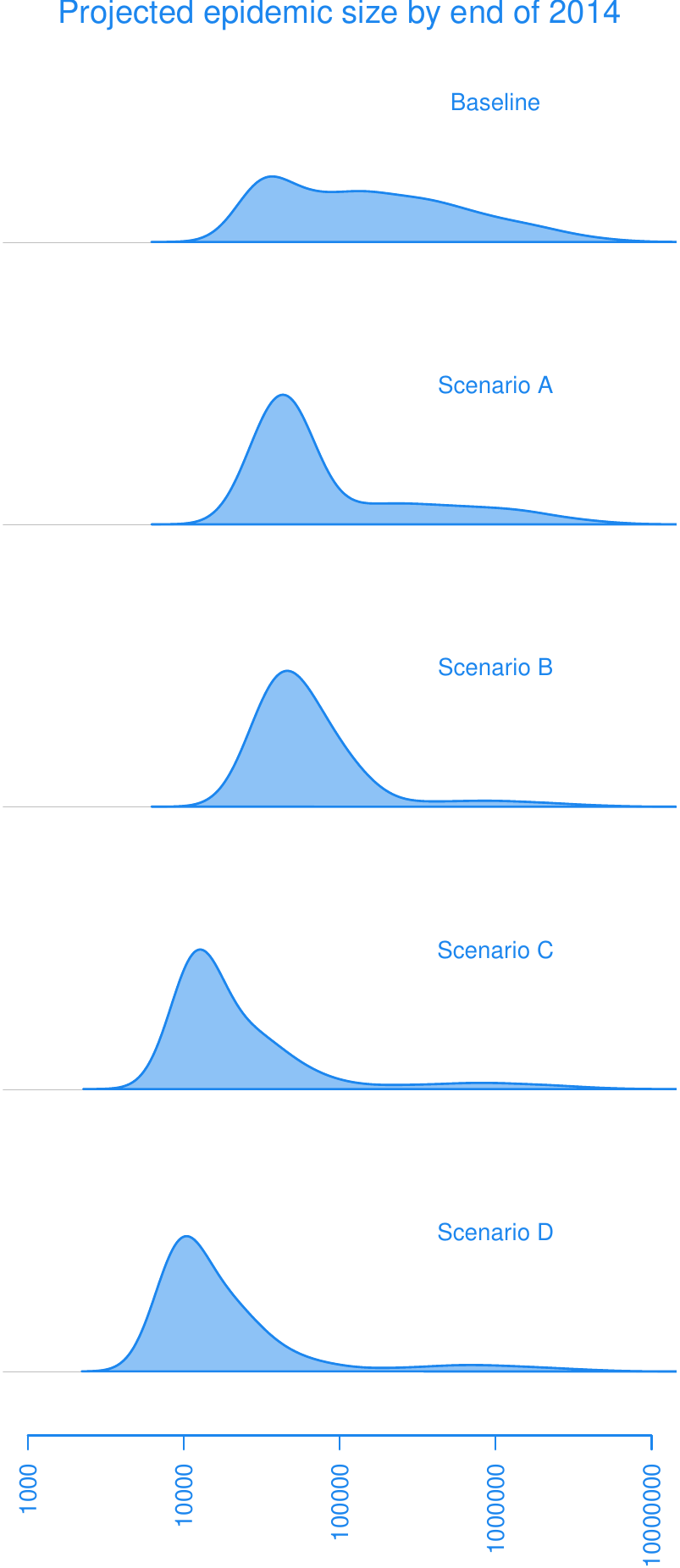}
\caption{Distribution of nuumber of cases by December 31, 2014 in five
scenarios. Scenario A reflects increased hospital capacity from US DoD
commitment of 15 September. Scenario B assumed significantly increased
hospital capacity in excess of Scenario A. Scenario C reflects
significantly increased hospital capacity and increased hospitalization
rates. Scenario D reflects significantly increased hospital capacity and
significantly increased hospitalization.}
\end{figure}

Scenario A assumes that the Department of Defense (DoD) improvements to
hospital capacity constitute the main intervention against the continued
spread of Ebola virus in West Africa. In this scenario, an additional
1,700 hospital beds become available between 25 October and 28 December
at a rate of one 100-bed facility every four days. This interpolation is
based on media reports that the first DoD unit is expected to come
online on 25 October and that all units are to be complete by the end of
2014. Results suggest that an initial downturn in cases is to be
expected based on isolation, but that capacity will be outstripped by
the continuing rise in cases. These results are consistent with
observations since 2 September (see online supplement). While these
results do predict a temporary downturn, they do not imply that
hospitalization is exclusively responsible for this trend. Particularly,
public compliance with burial policies, actions taken to increase
personal safety, and deterioration in reporting may all also play a
role. The median total projected epidemic size by 31 December is 51,202
cases (inter-quartile range: 37,868 to 152,453).

Scenario B assumes that the main line of intervention will be further
improvements to hospital capacity in excess of DoD improvements in
Scenario A. In this scenario, an additional 6,800 ``bed equivalents''
(which may include Ebola Community Care units as well as other units)
become available between 25 October and 28 December (including the 1,700
ETU beds from Scenario A). The outcome of this scenario is interesting
because it shows that improved treatment facilities are not enough to
ensure containment (see online supplement). As above, the increased
availability of treatment slows transmission for a time, but the
outbreak outgrows capacity and takes off again. The median total
projected epidemic size by 31 December is 51,260 cases (inter-quartile
range: 37,868 to 81,237). Thus, although the upper end of the
distribution is reduced, the median and lower end are similar to
Scenario A, suggesting that hospital capacity is unlikely to be the
limiting factor after the DoD improvements are complete.

Scenario C assumes that improved hospital capacity is complemented by
improved public compliance with recommendations. In this scenario, the
fraction of infected persons seeking hospitalization is increased from
its baseline (a variable number around 60\%) to 85\% in addition to the
increased hospital capacity envisioned in Scenario B. The expected
(median) outcome of this scenario is containment, although some
parameterizations cannot be contained by this strategy (see online
supplement). The median total projected epidemic size by 31 December is
14,829 cases (inter-quartile range: 11,608 to 30,192).

Although the expected (median) solution to Scenario C is rapid
containment with new cases peaking in mid-December, this is not
guaranteed. There are several plausible parameterizations for which the
interventions are inadequate. We therefore investigated a further
scenario (Scenario D) designed to achieve containment. In this scenario,
the number of additional bed equivalents is again 6,800 and public
compliance with hospitalization is 99\%. A majority of parameterizations
result in containment (see online supplement).

Since containment is achieved by a majority of parameterizations of this
scenario, we ran the simulation until elimination in a majority of
cases. The size and duration of 78\% of simulated epidemics ending by
mid-summer are summarized in the online supplement. Panels on the left
show the cumulative probability distribution for outbreak size (top) and
duration (bottom). The yellow band shows the inter-quartile range while
the median is indicated by a dashed line. Panels on the right show
histograms of outbreak size and duration from the $10,450$ simulations.
These results show that if the epidemic progresses as envisioned by
Scenario D, the total epidemic size is expected to be $\approx 12,285$
(inter-quartile range: 9,205 to 22,984) with elimination by mid-March.

\section{Discussion}\label{discussion}

The transmission of Ebola virus in West Africa continues to give rise to
high mortality and morbidity. Part of the challenge in predicting the
progression of the epidemic lies in the fundamentally different ways in
which transmission occurs: infection of hospital workers, community care
givers and those preparing bodies for funerals (World Health
Organization 2014d). Additionally, the timeframe and effectiveness of
increased hospital capacity compounds the problem of prediction
(Torjesen 2014), whether it is aimed at anticipating demand for
hospitalization or determining the level and speed of intervention
needed to bring the outbreak under control. Our approach was to
represent heterogeneity in transmission and time-varying intervention in
a multi-type branching process model (Jagers and Athreya 1997) that
offers analytic tractability, efficient simulation and the flexibility
to investigate a wide range of intervention scenarios. It is closely
related to sources of data; for example, stratifying cases into
hospital-treated versus community-treated allows for estimating
under-reporting which is thought to be large for the current epidemic
(Gomes et al. 2014).

Analytical insight, particularly the derivation of a reproductive ratio,
is useful when parameter estimates (such as the hospitalization rate)
are uncertain, since the sign and magnitude of their effects on
transmission can be derived. Besides recovering a full expression for
the basic reproductive ratio, simplifying assumptions such as assuming
that funeral-associated transmission can be reduced to zero, yield
further understanding. In particular, our model shows how the additional
exposure to health care workers in a hospital environment ($\beta$)
combines with both the reduced transmission in that environment
($\alpha$) and the hospitalization rate ($h$) to determine when
community (versus hospital) transmission will dominate ($i.e.$ when
$1-h>\alpha \beta$). Such formulas may provide ``rules of thumb'' to
help guide infection control or could improve practical decision making
by regularly updating estimates of core parameters through surveillance
within health facilities.

The approach we have taken to model parameterization is novel. A more
familiar approach is to propose a deterministic or stochastic model that
is then fit by minimizing an objective function on the errors, $e.g.$,
sum of squares or negative log likelihood of the data given the model
(Legrand et al. 2007). Statistical interpretation of such models (such
as hypothesis tests or confidence intervals) relies heavily on the
parametric specification of both the process model and the observation
model. If the proposed models are not good approximations to their
respective contributions to the data-generating process, then these
quantities may be quite biased. Moreover, such models are ineffective
when they are overparameterized. Our approach--the construction of
plausible parameter sets that are both epidemiologically sensible and
can reproduce observed properties of the epidemic--seeks to better
understand the space of models consistent with the data. The cost of
this approach is that the results do not admit probabilistic
interpretations, hypothesis tests, or the construction of confidence
intervals. A byproduct is that the identifiability of parameters (which
is compromised by overparameterization) is no longer an obstacle to
model construction and forecasting. If two parameters, say $a$ and $b$,
are highly correlated (not simultaneously identifiable) so that either
the model with large $a$ and small $b$ or large $b$ and small $a$ are
both consistent with the data, then the plausible set will include
parameter combinations with examples of both kinds (but not large $a$
and large $b$ or small $a$ and small $b$, say). It may be that these
differences are in fact irrelevant to the eventual behavior of the
model, in which case the space of possible solutions will be small.
Alternatively, it may be that these are just the parameters that most
substantially influence alternative outcomes, in which case the space of
possible solutions will be large. By seeking bounds on the range of
outcomes rather than a unique causal story, the method of plausible
parameter sets avoids technical problems with model identifiability and
more accurately emphasizes the kind of uncertainty prevalent under
emergent conditions while focusing attention on the property of most
practical interest: the possible future trajectories of the epidemic. In
conclusion, we believe that the method of plausible parameter sets is a
good starting point for exploring entire families of models and for
setting bounds on the range of possible outcomes. It is a first step
toward the construction of models for probabilistic inference.

In this study, we have focused on Liberia, one of the worst-hit
countries during the current outbreak. The ramping up of hospital
capacity in Liberia was dramatic during late August 2014, adding
approximately 300 beds. Throughout September, that sustained effort led
to an additional \textasciitilde{}300 beds. This heterogeneous increase
in capacity over time was incorporated into our model. We investigated
alternative hospital capacities and demands in a set of plausible
alternative scenarios. The best and worse outcomes of these scenarios
vary dramatically in the forecasted epidemic size. Median estimates are
at around 130,000 cases by 31 December 2014 assuming a baseline scenario
without increased hospital capacity. This is reduced to around 50,000
when capacity is ramped up to \textasciitilde{}1,700. Further increases
in hospital capacity reduce the upper bounds, but not the median. If the
hospitalization rate can be increased to 85\%, median predictions are of
containment, with an effective reproductive ratio \textless{}1. In
conclusion, this exercise suggests that in the absence of rapid
hospitalization of most cases, none of the proposed scenarios for
increasing hospital capacity is likely to achieve containment.

Branching process models use offspring distributions to simulate forward
in time. Here, the offspring of an infectious individual refers to the
new cases generated from that infectious individual. This is the type of
data that is frequently reported, even during early stages of an
outbreak. Models that require contact-tracing data are complicated by
the fact that there is uncertainty about whether contact is effective or
not. For example, how many ``contacts'' of an infectious individual that
gets on a plane are sufficiently intimate that infection is even a
causal possibility? Ambiguities about the causal relevance of contacts
of different kinds complicate models expressed in terms of attack rates.
By focusing on the empirical offspring distributions in various
transmission settings, one is able to build, simulate and analyze a
model with the key epidemiological features, and to investigate a wide
range of mitigation scenarios. In our case, the result was a multi-type
branching process that separated the location that infection was
acquired from the sites generating new infections. This approach
captures the behavioral aspects of transmission that are often lacking
in models (Funk, Salathé, and Jansen 2010). Awareness of Ebola in the
community and public education mean that community-acquired transmission
is increasingly likely to lead to demand for hospitalization. While our
methods are focused on the current Ebola outbreak in West Africa, they
apply to a broad class of infectious diseases.

\section{Acknowledgements}\label{acknowledgements}

We thank P. Rohani, J.P. Schmidt, and S. Scarpino for comments on an
earlier manuscript. Research reported here was supported by the National
Institute of General Medical Sciences of the National Institutes of
Health under Award Number U01GM110744. The content is solely the
responsibility of the authors and does not necessarily reflect the
official views of the National Institutes of Health.

\section*{References}\label{references}
\addcontentsline{toc}{section}{References}

Baize, Sylvain, Delphine Pannetier, Lisa Oestereich, Toni Rieger, Lamine
Koivogui, N'Faly Magassouba, Barrè Soropogui, et al. 2014. ``Emergence
of Zaire Ebola Virus Disease in Guinea.'' \emph{New England Journal of
Medicine} 0 (0): null.
doi:\href{http://dx.doi.org/10.1056/NEJMoa1404505}{10.1056/NEJMoa1404505}.
\url{http://www.nejm.org/doi/full/10.1056/NEJMoa1404505 http://www.nejm.org/doi/pdf/10.1056/NEJMoa1404505}.

Borchert, Matthias, Imaam Mutyaba, Maria D {Van Kerkhove}, Julius
Lutwama, Henry Luwaga, Geoffrey Bisoborwa, John Turyagaruka, et al.
2011. ``Ebola Haemorrhagic Fever Outbreak in Masindi District, Uganda:
outbreak Description and Lessons Learned.'' \emph{BMC Infectious
Diseases} 11 (1): 357.
doi:\href{http://dx.doi.org/10.1186/1471-2334-11-357}{10.1186/1471-2334-11-357}.
\url{http://www.biomedcentral.com/1471-2334/11/357}.

Briand, S., E. Bertherat, P. Cox, P. Formenty, K. Marie-Paule, J.K.
Myhre, C. Roth, Nahoko Shindo, and Christopher Dye. 2014. ``The
International Ebola Emergency.'' \emph{New Ewngland Journal of Medicine}
371: 1180--83. \url{http://www.nejm.org/doi/full/10.1056/NEJMp1409858}.

Bwaka, M A, M J Bonnet, P Calain, R Colebunders, A {De Roo}, Y Guimard,
K R Katwiki, et al. 1999. ``Ebola Hemorrhagic Fever in Kikwit,
Democratic Republic of the Congo: clinical Observations in 103
Patients.'' \emph{The Journal of Infectious Diseases} 179 Suppl
(Supplement\_1): S1--S7.
doi:\href{http://dx.doi.org/10.1086/514308}{10.1086/514308}.
\url{http://jid.oxfordjournals.org/content/179/Supplement/_1/S1.short}.

Chan, M. 2014. ``Ebola Virus Disease in West Africa --- No Early End to
the Outbreak.'' \emph{New England Journal of Medicine} 371: 1183--85.
\url{http://www.nejm.org/doi/full/10.1056/NEJMp1409859}.

Dowell, S F, R Mukunu, T G Ksiazek, A S Khan, P E Rollin, and C J
Peters. 1999. ``Transmission of Ebola Hemorrhagic Fever: a Study of Risk
Factors in Family Members, Kikwit, Democratic Republic of the Congo,
1995. Commission de Lutte Contre Les Epidémies À Kikwit.'' \emph{The
Journal of Infectious Diseases} 179 Suppl (Supplement\_1): S87--91.
doi:\href{http://dx.doi.org/10.1086/514284}{10.1086/514284}.
\url{http://jid.oxfordjournals.org/content/179/Supplement/_1/S87.short}.

Farrar, J.J., and P. Piot. 2014. ``The Ebola Emergency --- Immediate
Action, Ongoing Strategy.'' \emph{New England Journal of Medicine2}
3761: 1545--46. \url{http://www.nejm.org/doi/full/10.1056/NEJMe1411471}.

Fisman, David, Edwin Khoo, and Ashleigh Tuite. 2014. ``Early Epidemic
Dynamics of the West African 2014 Ebola Outbreak: Estimates Derived with
a Simple Two-Parameter Model.'' \emph{PLoS Currents}.
doi:\href{http://dx.doi.org/10.1371/currents.outbreaks.89c0d3783f36958d96ebbae97348d571}{10.1371/currents.outbreaks.89c0d3783f36958d96ebbae97348d571}.
\url{http://currents.plos.org/outbreaks/article/obk-14-0036-early-epidemic-dynamics-of-the-west-african-2014-ebola-outbreak-estimates-derived-with-a-simple-two-parameter-model/}.

Funk, Sebastian, Marcel Salathé, and Vincent A A Jansen. 2010.
``Modelling the Influence of Human Behaviour on the Spread of Infectious
Diseases: a Review.'' \emph{Journal of the Royal Society, Interface /
the Royal Society} 7 (50): 1247--56.
doi:\href{http://dx.doi.org/10.1098/rsif.2010.0142}{10.1098/rsif.2010.0142}.
\url{http://rsif.royalsocietypublishing.org/content/early/2010/05/25/rsif.2010.0142.full}.

Gomes, Marcelo F. C., Ana {Pastore y Piontti}, Luca Rossi, Dennis Chao,
Ira Longini, M. Elizabeth Halloran, and Alessandro Vespignani. 2014.
``Assessing the International Spreading Risk Associated with the 2014
West African Ebola Outbreak.'' \emph{PLoS Currents}.
doi:\href{http://dx.doi.org/10.1371/currents.outbreaks.cd818f63d40e24aef769dda7df9e0da5}{10.1371/currents.outbreaks.cd818f63d40e24aef769dda7df9e0da5}.
\url{http://currents.plos.org/outbreaks/?p=40803}.

Gulland, Anne. 2014. ``Spanish Authorities Investigate How Nurse
Contracted Ebola.'' \emph{BMJ (Clinical Research Ed.)} 349 (oct08\_6):
g6120.
doi:\href{http://dx.doi.org/10.1136/bmj.g6120}{10.1136/bmj.g6120}.
\url{http://www.bmj.com/content/349/bmj.g6120}.

Jagers, P., and Krishna B. Athreya. 1997. \emph{Classical and Modern
Branching Processes}. Springer.

Khan, A S, F K Tshioko, D L Heymann, B {Le Guenno}, P Nabeth, B
Kerstiëns, Y Fleerackers, et al. 1999. ``The Reemergence of Ebola
Hemorrhagic Fever, Democratic Republic of the Congo, 1995. Commission de
Lutte Contre Les Epidémies À Kikwit.'' \emph{The Journal of Infectious
Diseases} 179 Suppl (Supplement\_1): S76--86.
doi:\href{http://dx.doi.org/10.1086/514306}{10.1086/514306}.
\url{http://jid.oxfordjournals.org/content/179/Supplement/_1/S76.short}.

Legrand, J, R F Grais, P Y Boelle, A J Valleron, and A Flahault. 2007.
``Understanding the Dynamics of Ebola Epidemics.'' \emph{Epidemiology
and Infection} 135 (4). Cambridge University Press: 610--21.
doi:\href{http://dx.doi.org/10.1017/S0950268806007217}{10.1017/S0950268806007217}.
\url{http://journals.cambridge.org/abstract/_S0950268806007217}.

Medecins Sans Frontieres. 2014. ``Response to West Africa Ebola Epidemic
Remains Dangerously Inadequate \textbar{} MSF USA.''
\url{http://www.doctorswithoutborders.org/news-stories/field-news/response-west-africa-ebola-epidemic-remains-dangerously-inadequate}.

Meltzer, M.I., C. Y. Atkins, S. Santibanez, B. Knust, B.W. Petersen,
E.D. Ervin, S.T. Nichol, I.K. Damon, and M.L. Washington. 2014.
``Estimating the Future Number of Cases in the Ebola Epidemic ---
Liberia and Sierra Leone, 2014--2015.'' \emph{Morbidity and Mortality
Weekly Report} 63 (3): 1--1.
\url{http://www.cdc.gov/mmwr/preview/mmwrhtml/su6303a1.htm}.

Nishiura, H, and G Chowell. 2014. ``Early Transmission Dynamics of Ebola
Virus Disease (EVD), West Africa, March to August 2014.''
\emph{Eurosurveillance} 19 (36).
\url{http://www.eurosurveillance.org/ViewArticle.aspx?ArticleId=20894}.

Okware, S. I., F. G. Omaswa, S. Zaramba, A. Opio, J. J. Lutwama, J.
Kamugisha, E. B. Rwaguma, P. Kagwa, and M. Lamunu. 2002. ``An Outbreak
of Ebola in Uganda.'' \emph{Tropical Medicine and International Health}
7 (12): 1068--75.
doi:\href{http://dx.doi.org/10.1046/j.1365-3156.2002.00944.x}{10.1046/j.1365-3156.2002.00944.x}.
\url{http://doi.wiley.com/10.1046/j.1365-3156.2002.00944.x}.

Oyok, T. 2001. ``Outbreak of Ebola Hemorrhagic Fever Uganda, August
2000-January 2001.'' \emph{MMWR. Morbidity and Mortality Weekly Report}
50 (5): 73--77. \url{http://www.ncbi.nlm.nih.gov/pubmed/11686289}.

Roels, T H, A S Bloom, J Buffington, G L Muhungu, W R {Mac Kenzie}, A S
Khan, R Ndambi, et al. 1999. ``Ebola Hemorrhagic Fever, Kikwit,
Democratic Republic of the Congo, 1995: risk Factors for Patients
Without a Reported Exposure.'' \emph{The Journal of Infectious Diseases}
179 Suppl (Supplement\_1): S92--7.
doi:\href{http://dx.doi.org/10.1086/514286}{10.1086/514286}.
\url{http://jid.oxfordjournals.org/content/179/Supplement/_1/S92.short}.

Tomori, O, J Bertolli, P E Rollin, Y Fleerackers, Y Guimard, A {De Roo},
H Feldmann, et al. 1999. ``Serologic Survey Among Hospital and Health
Center Workers During the Ebola Hemorrhagic Fever Outbreak in Kikwit,
Democratic Republic of the Congo, 1995.'' \emph{The Journal of
Infectious Diseases} 179 Suppl (Supplement\_1): S98--S101.
doi:\href{http://dx.doi.org/10.1086/514307}{10.1086/514307}.
\url{http://jid.oxfordjournals.org/content/179/Supplement/_1/S98.short}.

Torjesen, Ingrid. 2014. ``World Leaders Are Ignoring Worldwide Threat of
Ebola, Says MSF.'' \emph{BMJ (Clinical Research Ed.)} 349 (sep05\_8):
g5496.
doi:\href{http://dx.doi.org/10.1136/bmj.g5496}{10.1136/bmj.g5496}.
\url{http://www.bmj.com/content/349/bmj.g5496}.

Towers, Sherry, Oscar Patterson-Lomba, and Carlos Castillo-Chavez. 2014.
``Temporal Variations in the Effective Reproduction Number of the 2014
West Africa Ebola Outbreak.'' \emph{PLoS Currents}.
doi:\href{http://dx.doi.org/10.1371/currents.outbreaks.9e4c4294ec8ce1adad283172b16bc908}{10.1371/currents.outbreaks.9e4c4294ec8ce1adad283172b16bc908}.
\url{http://currents.plos.org/outbreaks/?p=42655}.

UN-OCHA. 2014. ``Ebola Virus Outbreak - WEST AFRICA - April 2014.''
\url{http://fts.unocha.org/pageloader.aspx?page=emerg-emergencyDetails/\&emergID=16506}.

WHO Ebola Response Team. 2014. ``Ebola Virus Disease in West Africa ---
The First 9 Months of the Epidemic and Forward Projections.'' \emph{New
England Journal of Medicine}, September, 140922210513002.
doi:\href{http://dx.doi.org/10.1056/NEJMoa1411100}{10.1056/NEJMoa1411100}.
\url{http://www.nejm.org/doi/abs/10.1056/NEJMoa1411100}.

World Health Organization. 2014a. ``WHO Ebola Response Roadmap Situation
Report 22 October 2014.''

---------. 2014b. ``WHO \textbar{} Liberia: Ebola Treatment Centre Sets
a New Pace.'' World Health Organization.
\url{http://www.who.int/features/2014/liberia-ebola-island-clinic/en/}.

---------. 2014c. ``WHO \textbar{} Why the Ebola Outbreak Has Been
Underestimated.'' World Health Organization.
\url{http://www.who.int/mediacentre/news/ebola/22-august-2014/en/}.

---------. 2014d. ``Ebola Virus Disease in West Africa --- The First 9
Months of the Epidemic and Forward Projections.'' \emph{New England
Journal of Medicine} 371: 1481--95.
\url{http://www.nejm.org/doi/full/10.1056/NEJMoa1411100}.

---------. 2014e. ``Ebola Virus Disease, Liberia (Situation as of 30
March 2014) - WHO \textbar{} Regional Office for Africa.''
\url{http://www.afro.who.int/en/clusters-a-programmes/dpc/epidemic-a-pandemic-alert-and-response/outbreak-news/4072-ebola-virus-disease-liberia.html}.

---------. 2014f. ``Ebola Virus Disease, West Africa (Update of 26 May
2014) - WHO \textbar{} Regional Office for Africa.''
\url{http://www.afro.who.int/en/clusters-a-programmes/dpc/epidemic-a-pandemic-alert-and-response/outbreak-news/4143-ebola-virus-disease-west-africa-26-may-2014.html}.

---------. 2014g. ``Ebola Virus Disease, West Africa -- Update 25 July
2014 - WHO \textbar{} Regional Office for Africa.''
\url{http://www.afro.who.int/en/clusters-a-programmes/dpc/epidemic-a-pandemic-alert-and-response/outbreak-news/4233-ebola-virus-disease-west-africa-25-july-2014.html}.

---------. 2014h. ``WHO \textbar{} Ebola Virus Disease Update -
Senegal.'' \url{http://www.who.int/csr/don/2014/_08/_30/_ebola/en/}.

---------. 2014i. ``WHO \textbar{} WHO Statement on the Meeting of the
International Health Regulations Emergency Committee Regarding the 2014
Ebola Outbreak in West Africa.''
\url{http://www.who.int/mediacentre/news/statements/2014/ebola-20140808/en/}.

---------. 2014j. ``WHO \textbar{} Ebola Virus Disease -- United States
of America.''
\url{http://www.who.int/csr/don/01-october-2014-ebola/en/}.

\end{document}

\maketitle

\section{Introduction}\label{introduction}

This document contains supplementary material for the paper ``Ebola
cases and health system demand in Liberia'' by the UGA-MIDAS Ebola
Modeling Group.

\section{Effective reproduction number
($R_{eff}$)}\label{effective-reproduction-number-rux5feff}

From the independence of the mixture components, we obtain the
\emph{mean matrix} for transmission generations defined in terms of
treatment location.

\begin{equation}
  M =
  \begin{bmatrix}
    N q \alpha \beta \theta + h \lambda_h & (1-h) \lambda_h \\
    h (\theta N q g + (1-g)(N q \theta + \phi)) & (1-h)(\theta N q g + (1-g)(N q \theta + \phi))
  \end{bmatrix}
\end{equation}

The two eigenvalues of this matrix are:

\begin{multline}
  \Lambda_1 = \frac{1}{2}(Nq(1-h+\alpha \beta)\theta + (g-1)(h-1)\phi + h \lambda_h \\ + \sqrt{((Nq\phi(h+\alpha \beta -1)-(g-1)(h-1)\phi)^2 + h \lambda_h(2Nq\phi(1-h+\alpha \beta)+2\phi(g-1)(h-1)+h\lambda_h))})
\end{multline}

\begin{multline}
  \Lambda_2 = \frac{1}{2}(Nq(1-h+\alpha \beta)\theta + (g-1)(h-1)\phi + h \lambda_h \\ - \sqrt{((Nq\phi(h+\alpha \beta -1)-(g-1)(h-1)\phi)^2 + h \lambda_h(2Nq\phi(1-h+\alpha \beta)+2\phi(g-1)(h-1)+h\lambda_h))})
\end{multline}

The dominant eigenvalue ($\Lambda$) is the long run growth rate of the
epidemic and provides a threshold criterion such that outbreak will grow
if $\Lambda > 1$ and decline if $\Lambda < 1$. In this model, which
ignores susceptible depletion, $\Lambda$ is always the effective
reproduction number ($R_{eff}$) in that it is the average number of
secondary infections in a population comprised of community-treated and
hospital-treated cases at its stable distribution. If evaluated at
$t=0$, $\Lambda$ may also be interpreted as the basic reproductive ratio
($R_0$). A special case of interest is the complete elimination of cases
in the community generated by cases treated in the hospital
($\lambda_h=0$). In this case, the eigenvalues are
$\Lambda_1 = \alpha \beta \theta N q$ and
$\Lambda_2 = (1-h)((1-g)(\theta Nq + \phi) + g \theta N q)$. Which
$\Lambda$ will be dominant depends on the values of $\alpha$, $\beta$,
$h$, $g$, and $\phi$, so that eventually either community transmission
or hospital transmission drives the persistence of the infection.
Further insight may be obtained by inspecting the case where funeral
transmission is reduced to zero ($\phi=0$). Then,
$\Lambda_2 = (1-h)\theta N q$. Community transmission dominates in this
case if $1-h > \alpha \beta$. Note that where hospital transmission
dominates ($\Lambda_1 > \Lambda_2$) the elasticities of the parameters
are identical. This means that proportional changes in each quantity
have identical effect (halving the contact number is equivalent to
halving the effectiveness of infection control is equivalent to halving
the increased contact rate in health care facilities, \emph{etc.}).

\section{Supplementary Figures}\label{supplementary-figures}

Figure 1 shows the distribution of five case classifications over the
time period used for fitting. These are: (1) number of health care
workers infected (HCW), (2) total number of reported cases, (3) total
cases, (4) fraction of cases that were hospital-acquired, and (5)
fraction of cases associated with funeral preparation and burial. For
(1) and (2), the red line corresponds to the WHO reported number during
the interval. For (3), the red line corresponds to total cases obtained
by multiplying the number of reported cases by 2.5, the presumed factor
of under-reporting. Importantly, the number of funeral acquired cases is
consistent with anecdotes reported by Rivers et al (2014). The fraction
of cases that are hospital-acquired is somewhat lower than anecdotally
reported. Figure 2 projects daily hospital demand (new patients seeking
hospitalization) according to the baseline scenario. Figures 3 and 4
illustrate the range of trajectories and daily hospital demand
associated with Scenario A. Figures 5 and 6 illustrate the range of
trajectories and daily hospital demand associated with Scenario B.
Figures 7 and 8 illustrate the range of trajectories and daily hospital
demand associated with Scenario C. Figures 9 and 10 illustrate the range
of trajectories and daily hospital demand associated with Scenario D.
Figure 11 shows the total epidemic duration and outbreak size for the
78\% of simulations of Scenario D terminating in mid summer 2015.

\begin{figure}[htbp]
\centering
\includegraphics{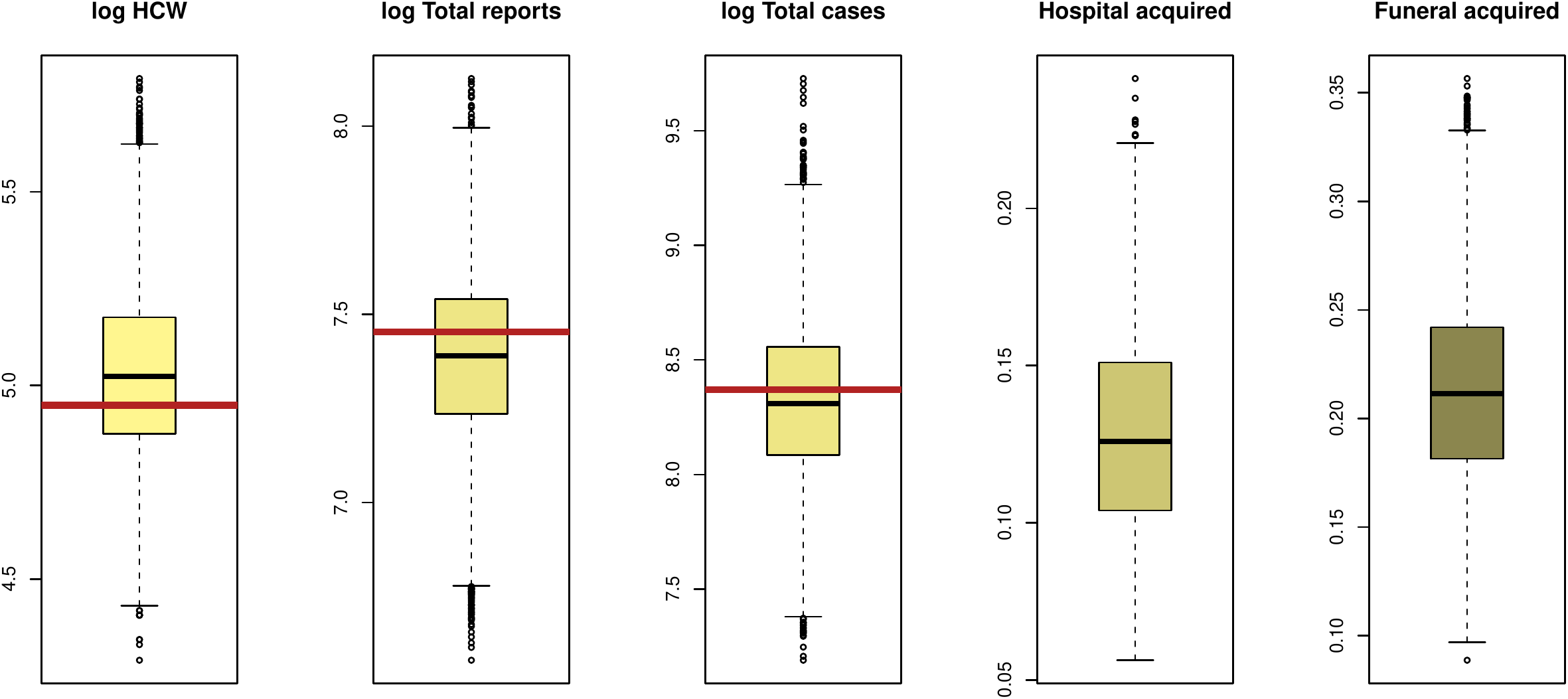}
\caption{Cumulative reported cases in data (red lines) and model
simulations (box-and-whisker plots). The left three panels show results
for health care workers, reported cases, and reported and unreported
cases (assuming 2.5 fold under-reporting). The remaining panels show the
model-predicted distributions of hospital-acquired infections and
funeral-acquired infections.}
\end{figure}

\begin{figure}[htbp]
\centering
\includegraphics{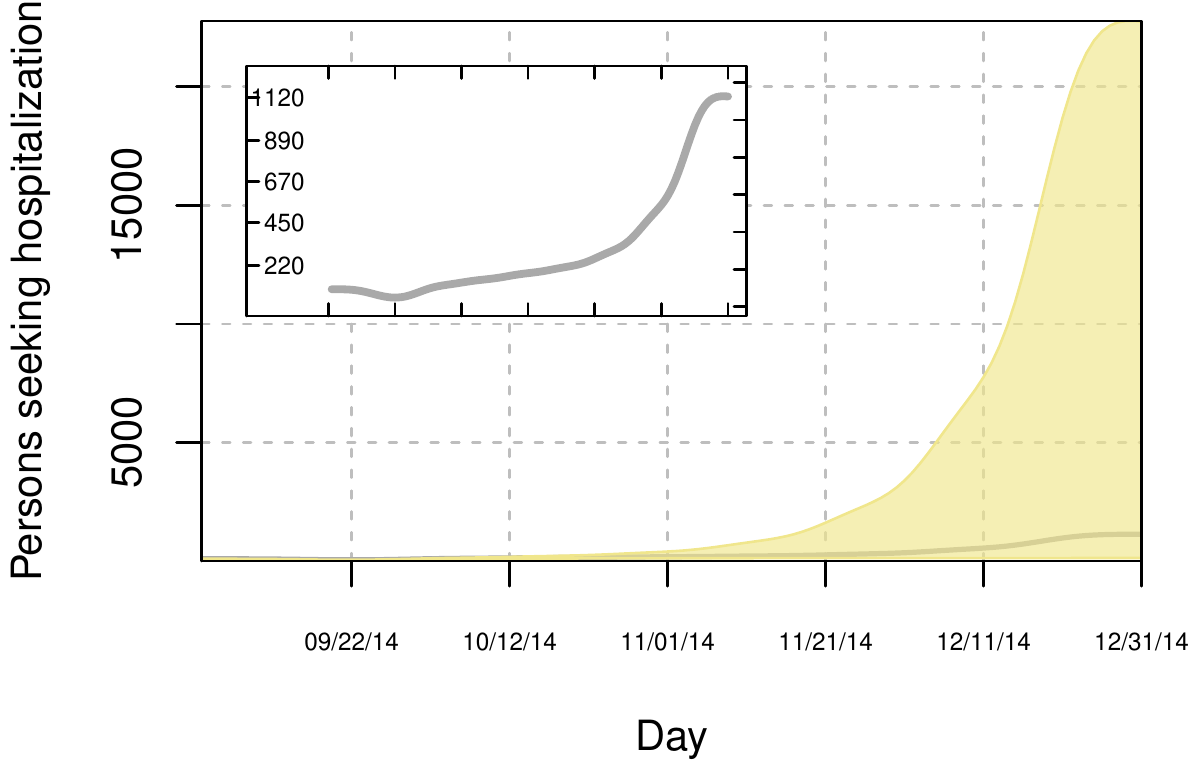}
\caption{Daily number seeking hospitalization (bottom) when transmission
occurs at the baseline rate. Inset plot shows the median daily number of
cases.}
\end{figure}

\begin{figure}[htbp]
\centering
\includegraphics{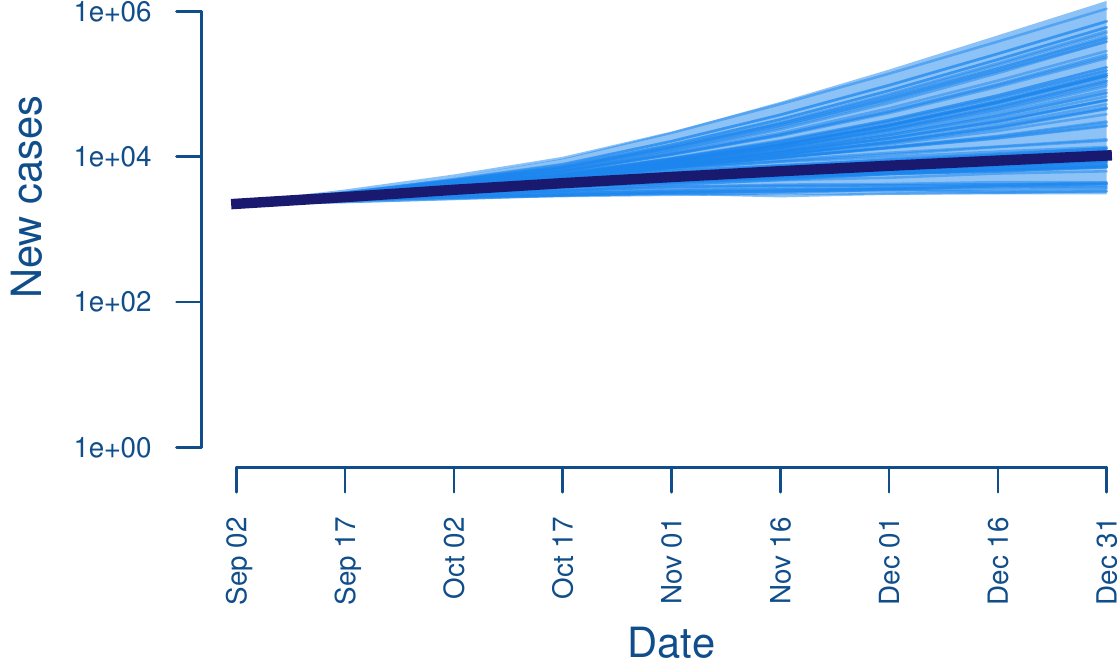}
\caption{Number of cases in each infection generation under Scenario A.}
\end{figure}

\begin{figure}[htbp]
\centering
\includegraphics{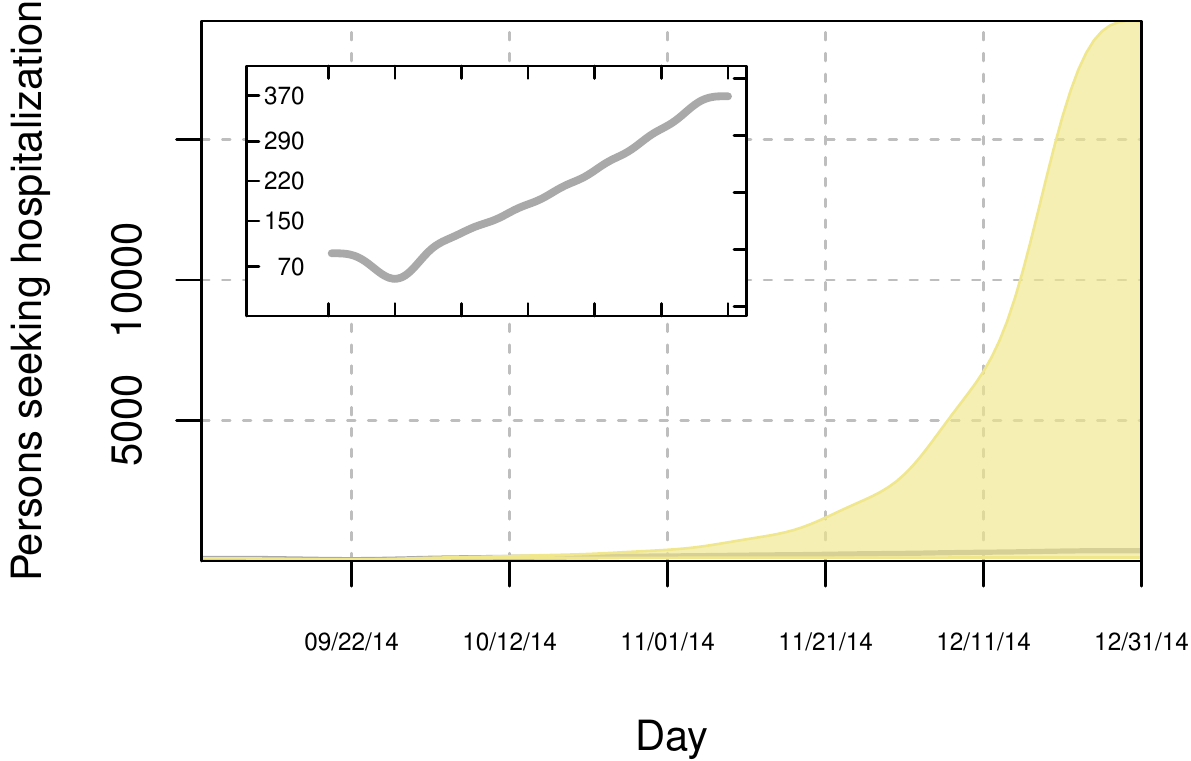}
\caption{Daily number seeking hospitalization (bottom) according to
Scenario A. Inset plot shows the median daily number of cases.}
\end{figure}

\begin{figure}[htbp]
\centering
\includegraphics{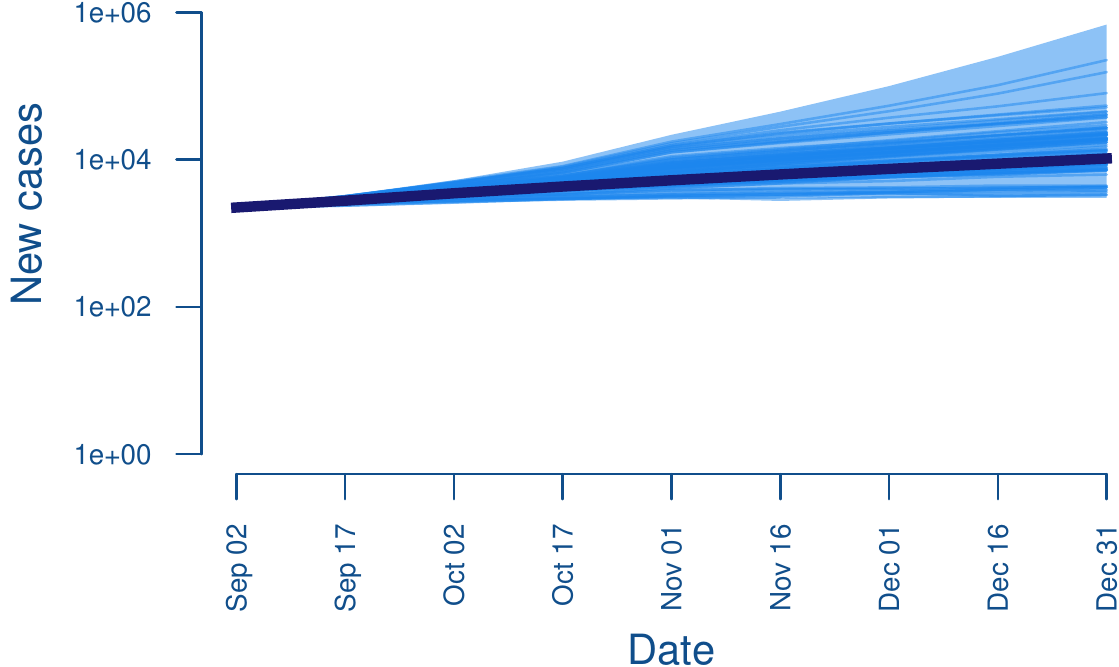}
\caption{Number of cases in each infection generation under Scenario B.}
\end{figure}

\begin{figure}[htbp]
\centering
\includegraphics{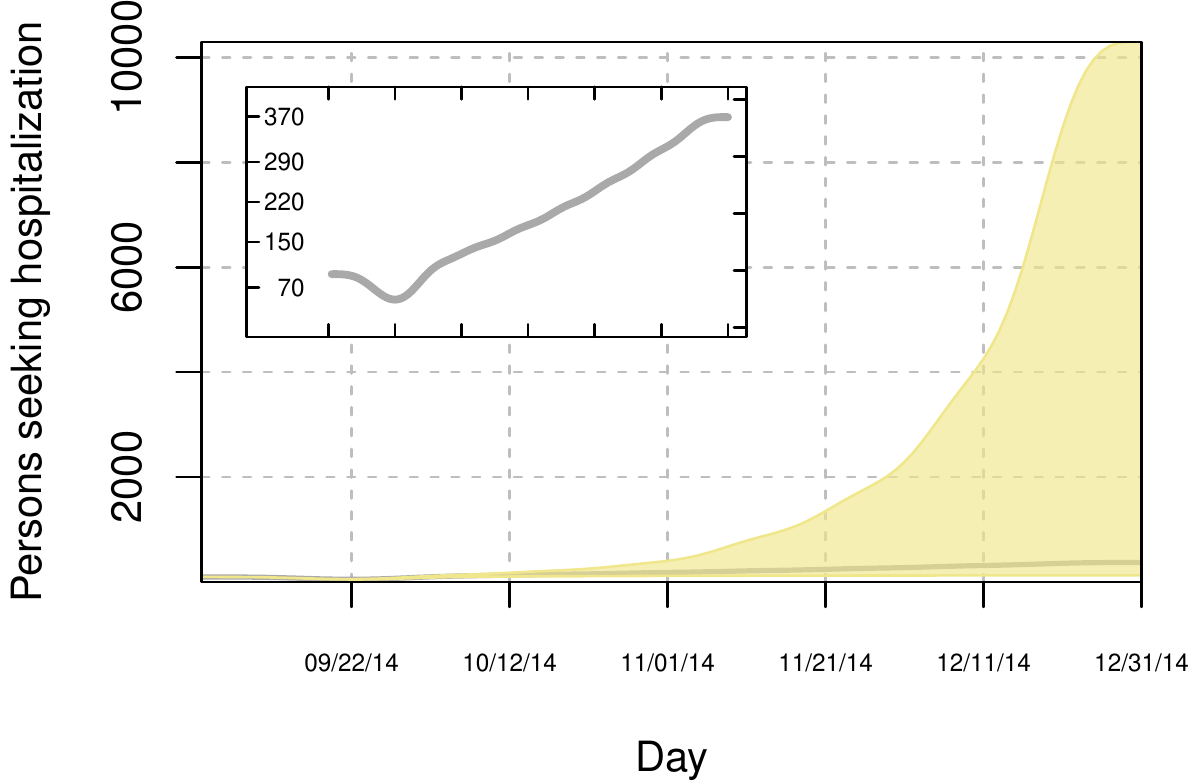}
\caption{Daily number seeking hospitalization (bottom) according to
Scenario B. Inset plot shows the median daily number of cases.}
\end{figure}

\begin{figure}[htbp]
\centering
\includegraphics{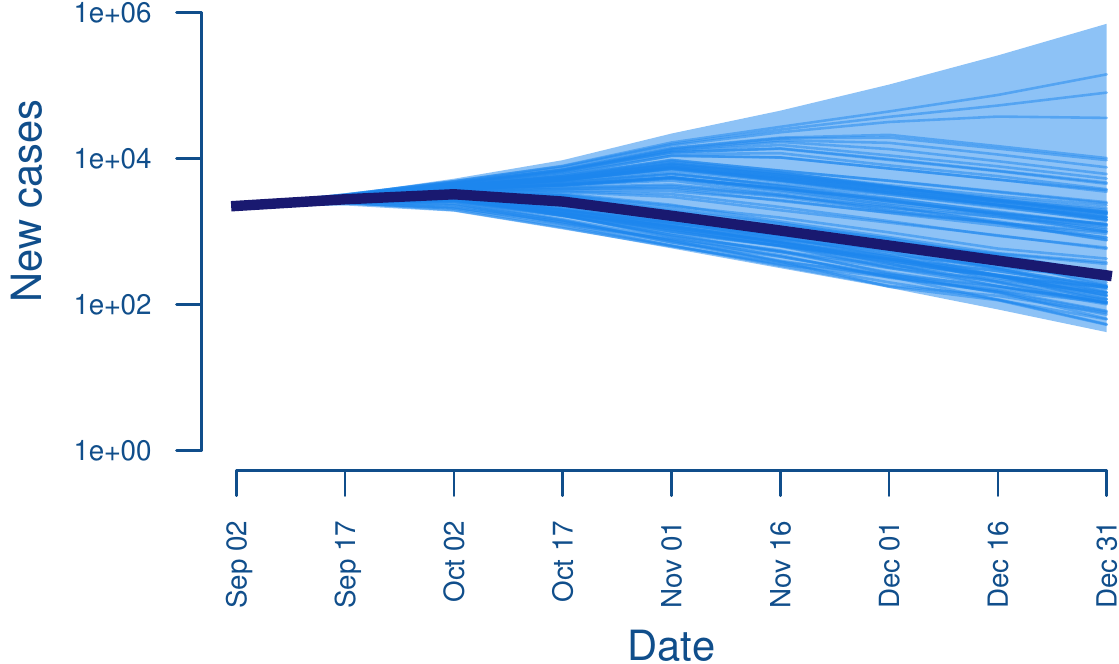}
\caption{Number of cases in each infection generation under Scenario C.}
\end{figure}

\begin{figure}[htbp]
\centering
\includegraphics{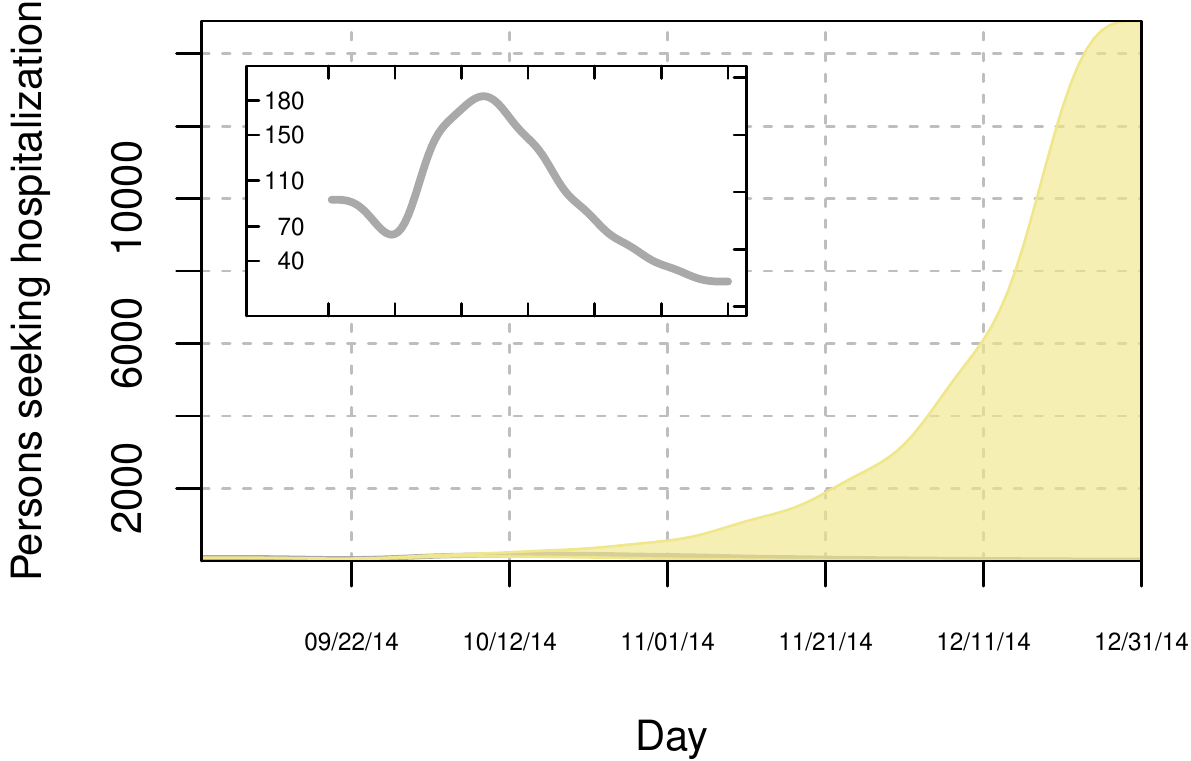}
\caption{Daily number seeking hospitalization (bottom) according to
under Scenario C. Inset plot shows the median daily number of cases.}
\end{figure}

\begin{figure}[htbp]
\centering
\includegraphics{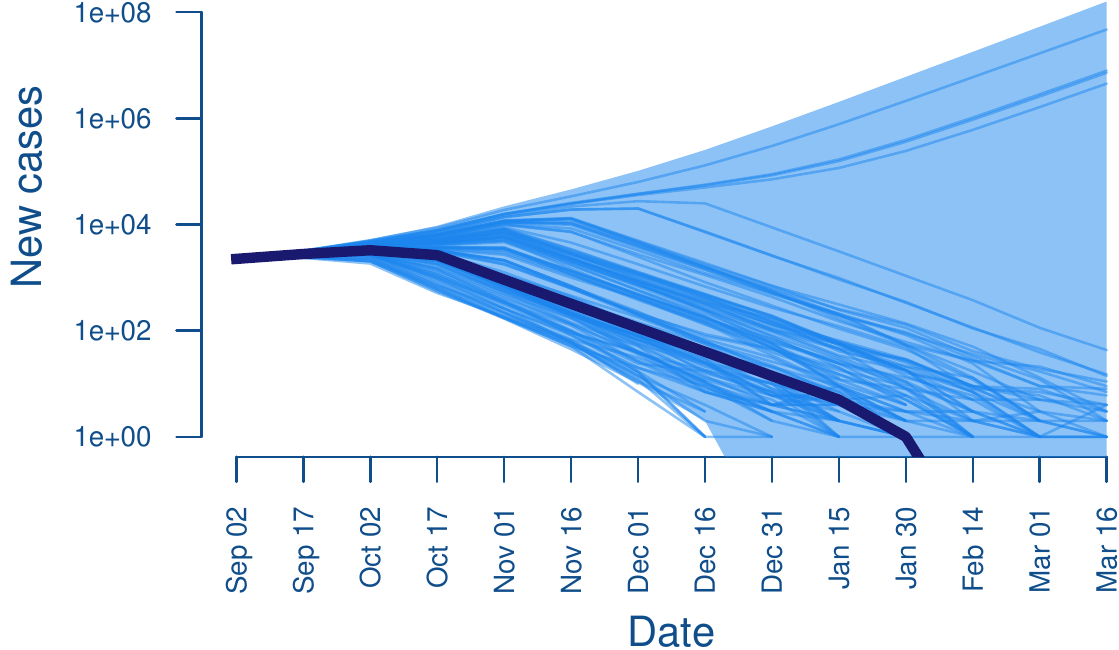}
\caption{Number of cases in each infection generation under Scenario D.}
\end{figure}

\begin{figure}[htbp]
\centering
\includegraphics{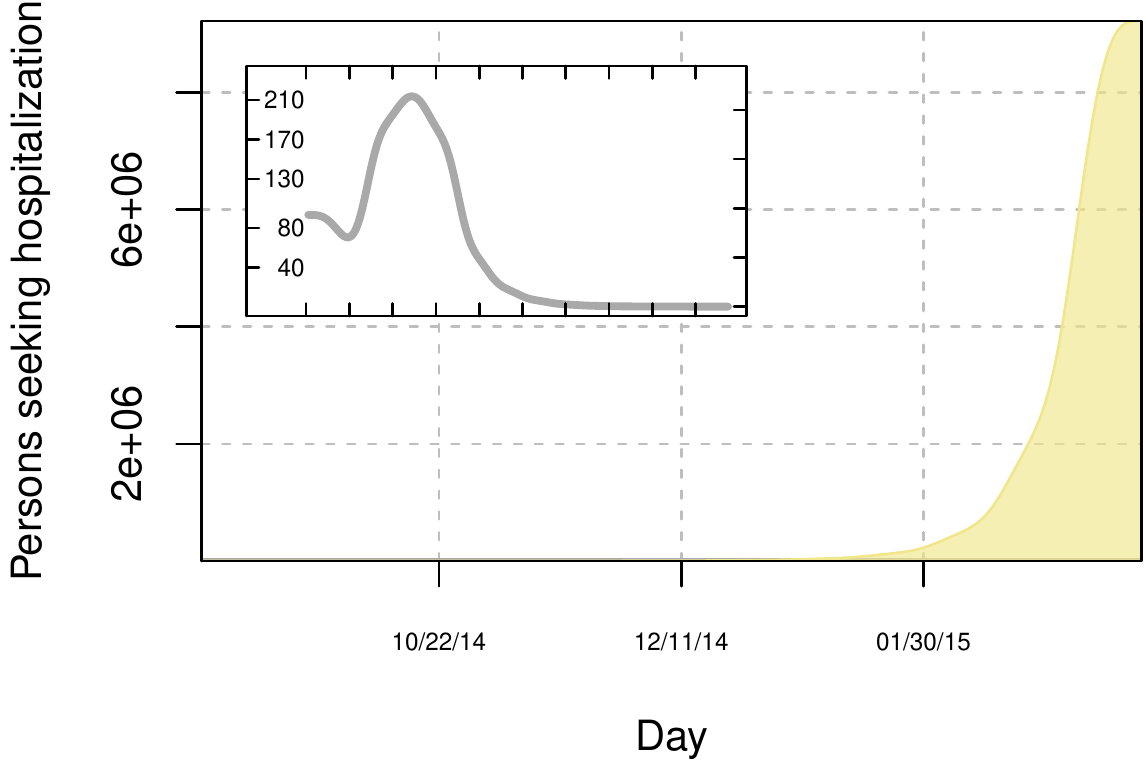}
\caption{Daily number seeking hospitalization (bottom) according to
under Scenario D. Inset plot shows the median daily number of cases.}
\end{figure}

\begin{figure}[htbp]
\centering
\includegraphics{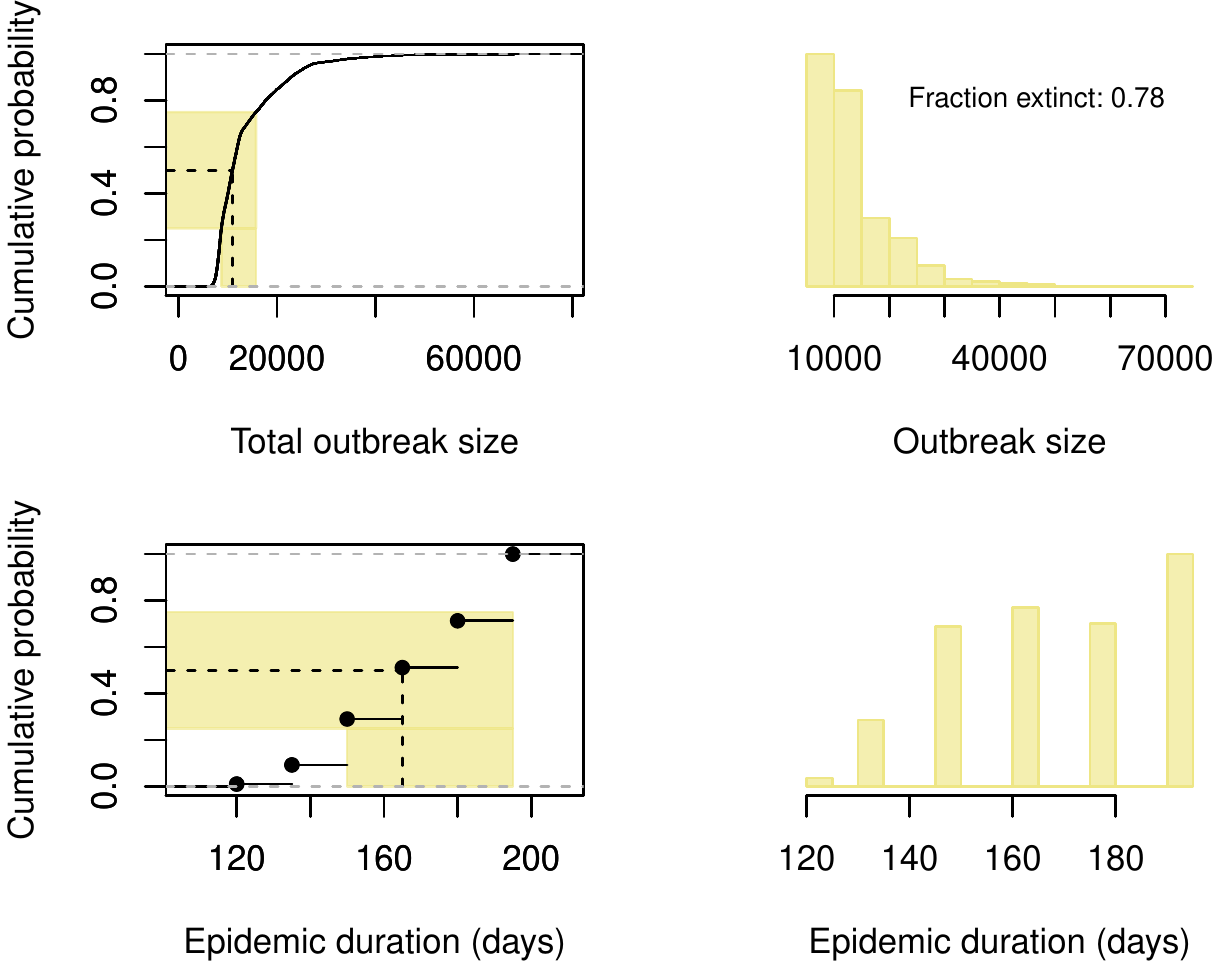}
\caption{Cumulative distribution function (left) and histogram (right)
of the total epidemic size (top) and epidemic duration in days after 2
September 2014 (bottom) in a containment scenario (Scenario C).}
\end{figure}